\documentclass[twocolumn]{aastex63}

\usepackage{enumerate}
\usepackage{graphicx}
\usepackage{longtable,booktabs}
\usepackage{multirow}
\usepackage{appendix}
\usepackage{float}
\usepackage{amssymb}
\usepackage{lineno}
\usepackage{threeparttable}
\usepackage{soul}
\usepackage{needspace}
\submitjournal{APJ}

\accepted{Nov. 1, 2022}

\shorttitle{The H-poor Superluminous Supernovae from ZTF-I}
\shortauthors{Chen et al.}
\watermark{Draft}
\graphicspath{{./}{figures/}}
\begin{document}

%\title{The Hydrogen-Poor Superluminous Supernovae from the Zwicky Transient Facility Phase-I Survey: II. Light Curve Modeling and Analysis}
\title{The Hydrogen-Poor Superluminous Supernovae from the Zwicky Transient Facility Phase-I Survey: II. Light Curve Modeling and Characterization of Undulations}

\author[0000-0001-5175-4652]{Z.~H.~Chen}
\affil{Physics Department and Tsinghua Center for Astrophysics (THCA), Tsinghua University, Beijing, 100084, China}

\author[0000-0003-1710-9339]{Lin~Yan}
\affil{Caltech Optical Observatories, California Institute of Technology, Pasadena, CA 91125, USA}

\author[0000-0002-5477-0217]{T.~Kangas}
\affil{The Oskar Klein Centre, KTH Royal Institute of Technology, Stockholm, Sweden}

\author[0000-0001-9454-4639]{R.~Lunnan}
\affil{The Oskar Klein Centre, Department of Astronomy, Stockholm University, AlbaNova, SE-106 91 Stockholm, Sweden}

\author[0000-0003-1546-6615]{J.~Sollerman}
\affil{The Oskar Klein Centre, Department of Astronomy, Stockholm University, AlbaNova, SE-106 91 Stockholm, Sweden}

\author{S.~Schulze}
\affil{The Oskar Klein Centre, Department of Physics, Stockholm University, AlbaNova, SE-106 91 Stockholm, Sweden}

\author{D.~A.~Perley}
\affil{Astrophysics Research Institute, Liverpool John Moores University, 146 Brownlow Hill, Liverpool L3 5RF, UK}

\author[0000-0002-1066-6098]{T.-W.~Chen}
\affil{The Oskar Klein Centre, Department of Astronomy, Stockholm University, AlbaNova, SE-106 91 Stockholm, Sweden}

\author{K.~Taggart}
\affil{Department of Astronomy and Astrophysics, University of California, Santa Cruz, CA 95064, USA}

\author{K.~R.~Hinds}
\affiliation{Astrophysics Research Institute, Liverpool John Moores University, 146 Brownlow Hill, Liverpool L3 5RF, UK}

\author{A.~Gal-Yam}
\affil{Department of particle physics and astrophysics, Weizmann Institute of Science, 76100 Rehovot, Israel}

\author[0000-0002-7334-2357]{X.~F.~Wang}
\affil{Physics Department and Tsinghua Center for Astrophysics (THCA), Tsinghua University, Beijing, 100084, China}
\affil{Beijing Planetarium, Beijing Academy of Sciences and Technology, Beijing, 100044, China}

\author{K.~De}
\affil{Cahill Center for Astrophysics, California Institute of Technology, 1200 E. California Blvd. Pasadena, CA 91125, USA}

%\author{+builderlist}
\author[0000-0001-8018-5348]{E.~Bellm}
\affiliation{DIRAC Institute, Department of Astronomy, University of Washington, 3910 15th Avenue NE, Seattle, WA 98195, USA}
\author[0000-0002-7777-216X]{J.~S.~Bloom}
\affiliation{Department of Astrophysics, University of California, Berkeley, CA 94720-3411, USA}
\affiliation{Lawrence Berkeley National Laboratory, 1 Cyclotron Road, MS 50B-4206, Berkeley, CA 94720, USA}
\author[0000-0002-5884-7867]{R.~Dekany}
\affil{Caltech Optical Observatories, California Institute of Technology, Pasadena, CA 91125, USA}
%\author{C.~Fremling}
%\affil{Division of Physics, Mathematics, and Astronomy, California Institute of Technology, Pasadena, CA 91125, USA}
\author{M.~Graham}
\affil{Cahill Center for Astrophysics, California Institute of Technology, 1200 E. California Blvd. Pasadena, CA 91125, USA}
\author{M.~Kasliwal}
\affil{Division of Physics, Mathematics, and Astronomy, California Institute of Technology, Pasadena, CA 91125, USA}
\author{S.~Kulkarni}
\affil{Division of Physics, Mathematics, and Astronomy, California Institute of Technology, Pasadena, CA 91125, USA}
\author[0000-0003-2451-5482]{R.~Laher}
\affiliation{IPAC, California Institute of Technology, 1200 E. California Blvd, Pasadena, CA 91125, USA}
\author{D.~Neill}
\affil{Cahill Center for Astrophysics, California Institute of Technology, 1200 E. California Blvd. Pasadena, CA 91125, USA}
\author[0000-0001-7648-4142]{B.~Rusholme}
\affiliation{IPAC, California Institute of Technology, 1200 E. California Blvd, Pasadena, CA 91125, USA}

\correspondingauthor{Lin~Yan, Z.~H.~Chen}
\email{lyan@caltech.edu, chenzh18@mails.tsinghua.edu.cn}

%% Mark off the abstract in the ``abstract'' environment. 
\begin{abstract}
We present analysis of the light curves (LCs) of 77 hydrogen-poor superluminous supernovae (SLSNe-I) discovered during the Zwicky Transient Facility Phase-I operation. We find that the majority (67\%) of the sample can be fit equally well by both magnetar and ejecta-circumstellar medium (CSM) interaction plus $^{56}$Ni decay models. This implies that LCs alone can not unambiguously constrain the physical power sources for a SLSN-I. However, 23\% of the sample show inverted V-shape, steep declining LCs or features of long rise and fast post-peak decay, which are better described by the CSM+Ni model. The remaining 10\% of the sample favor the magnetar model. Moreover, our analysis shows that the LC undulations are quite common, with a fraction of $18-44\% $ in our gold sample. Among those strongly undulating events, about 62\% of them are found to be CSM-favored, implying that the undulations tend to occur in the CSM-favored events. Undulations show a wide range in energy and duration, with median values (and 1$\sigma$ errors) being as $1.7\%^{+1.5\%}_{-0.7\%}\,\rm E_{\rm rad,total}$ and $28.8^{+14.4}_{-9.1}$\,days, respectively. Our analysis of the undulation time scales suggests that intrinsic temporal variations of the central engine can explain half of the undulating events, while CSM interaction can account for the majority of the sample. Finally, all of the well-observed He-rich SLSNe-Ib have either strongly undulating LCs or the LCs are much better fit by the CSM+Ni model. These observations imply that their progenitor stars have not had enough time to lose all of the He-envelopes before supernova explosions, and H-poor CSM are likely to present in these events. 
\end{abstract}

\keywords{Stars: supernovae: general}

\section{Introduction} \label{sec:intro}

Superluminous supernovae (SLSNe), as one group of energetic stellar explosions, were first discovered in the mid-2000s \citep{Quimby2007,Ofek2007,Smith2007}. They are 10 -- 100 times more luminous at the peak phase and evolve much slower than normal Type Ia and core-collapse supernovae (SNe). After the initial discoveries, it quickly became clear that the conventional radioactive decay model for normal core-collapse supernovae (CCSNe) can not explain the majority of SLSNe. Today, what powers these luminous and slowly evolving events remains unclear. Several mechanisms have been proposed, including interaction with circumstellar material \citep[CSM,][]{Chevalier2011,Chatzopoulos2012,Chatzopoulos2013,Benetti_2014}, energy injection from a central engine such as a rapidly rotating neutron star \citep[magnetar,][]{Kasen2010,Woosley2010} or an accreting black hole \citep{Dexter2013}. Exotic, rare explosions were also proposed, such as electron-positron pair instability or pulsational pair instability supernovae theoretically predicted for extremely massive stars \citep[PISN or PPISN,][]{Rakavy1967,Barkat1967,Woosley2007,Woosley2017}. 

Magnetar models are flexible and often used to fit the light curves of SLSNe-I \citep[{\textit{e.g.}}][]{Inserra_2013,Nicholl2013}. However, some observations already indicate that magnetar spin-down is not the only process which affects the LC luminosity and morphology for a SLSN-I, and there might be multiple processes affecting the optical emission. For example, the detection of late time H$\alpha$ emission in three SLSNe-I indicates the presence of H-rich CSM shells ejected by the progenitor stars \citep{Yan2015,Yan2017a}. Another example is the discovery of Mg\,II emission lines resonant scattered by a H-poor CSM shell in the SLSN-I PTF16eh \citep{Lunnan2018b}.  Finally, the sharp V-shaped LCs of SN\,2017egm are shown to be better fit by the ejecta-CSM interaction (CSI) model \citep{Wheeler2017}.  Statistically, it is not clear what roles magnetars and CSI play for the population of SLSNe-I, and this needs further studies. 

Temporal bumps or dips in SLSN-I LCs are known and have been observed in various objects \citep{Nicholl_2016b,Yan2017a,Yan2017b,Anderson2018,Lunnan2020}. LCs with such undulations can not be explained by a simple magnetar model or CSI model. Previously, poor LC sampling and lack of uniform SLSN-I data sets have precluded detailed statistical analysis of LC undulations. A recent paper by \citet[][cited as \hyperlink{cite.Hosseinzadeh2022}{H22} hereafter]{Hosseinzadeh2022} has carried out a focused study using the published SLSN-I LCs from the literature. It discussed various possible physical drivers for this phenomenon but found no conclusive answers. The \hyperlink{cite.Hosseinzadeh2022}{H22} sample is compiled from multiple sources, which can introduce biases to undulation fractions and properties. The Zwicky Transient Facility (ZTF) Phase-I survey \citep{Graham2019,Bellm2019,Masci2019} has discovered and classified a large sample of SLSNe-I. The advantages of the ZTF LCs are the high cadence (3 days or less) and the excellent phase coverage at both early and late times \citep{Bellm2019b}. This provides a valuable opportunity to perform a statistical study on the LC undulations.

The description of the ZTF survey, the SLSN-I sample and the complete photometric dataset are published in \citet{Chen2022}, Paper I of this series.  Paper I presented mostly the parameters which can be measured or directly computed from the data, such as redshift, extinction correction, K-correction, peak luminosity, peak phase, time scales (rise \&\ decline), color, black-body temperature and bolometric luminosity. 

%This paper, Paper II, presents the detailed analysis of the LCs of the 77 of 78 ZTF SLSNe-I published in Paper I, excluding SN\,2018ibb which is the focus of a separate paper (Schulze et al. in preparation). 
%This paper, Paper II, presents the detailed analysis of the LCs of the 77 ZTF SLSNe-I published in Paper I (SN\,2018ibb is not included for the lack of photometry data). 

This paper, Paper II, presents the detailed analysis of the LCs of the 77 of 78 ZTF SLSNe-I published in Paper I, excluding SN\,2018ibb. The LC and detailed analysis of SN\,2018ibb will be published in Schulze et al. (in preparation). This is a well-sampled events with evident undulations and excluding it lower our undulation fraction for about 2\%. 
We focus on the LC morphology and various physical parameters derived from modeling, such as ejecta mass and explosion energy. Throughout the paper, apparent magnitudes are in the AB system, unless specified otherwise. 
We adopt a $\Lambda$CDM cosmology with H$_0 = 70.0$\,km\,s$^{-1}$\,Mpc$^{-1}$, $\Omega_M = 0.3$ and $\Omega_{\Lambda} = 0.7$. 
%We adopt a $\Lambda$CDM cosmology with H$_0 = 67.7$\,km\,s$^{-1}$\,Mpc$^{-1}$, $\Omega_M = 0.307$ and $\Omega_{\Lambda} = 0.693$ \citep{Planck2016}. 

\section{The Data}
\label{sec:data}
Our sample contains 77 SLSNe-I discovered from March 17, 2018 to October 31, 2020 by the ZTF survey. This sample covers redshifts of $z \sim 0.06 - 0.67$. The photometry data primarily comes from the ZTF in the $g,r,i$ bands \citep{Bellm2019}, and also includes additional data from other ground-based facilities (see Paper I for details) and {\it Swift} \citep{Roming2005}. 
%Note that the systematic offsets between different telescopes have been evaluated and combined in the errors (see \S~4.1.3 of Paper I).
Each event has been spectroscopically classified as described in Paper I. The majority of the spectra used for the velocity measurements are from the Double Beam Spectrograph \citep[DBSP,][]{Oke1983} and the Low Resolution Imaging Spectrometer \citep[LRIS,][]{Oke1995} mounted on the Palomar 200\,inch (P200) and the Keck telescope respectively. The spectra used for velocity measurement but not presented in Paper I will be published in our future work.

We divide the ZTF SLSN-I sample into three subsets -- `gold', `silver', and `bronze'.  %This is because some results depend on the LC phase coverage. 
This is because the undulations can be missed without sufficiently sampled data, and some results ({\it e.g.} the fraction of undulating events in \S~\ref{subsec:Undulation}) strongly depend on the LC phase coverage. 
%The bronze class is defined by the number of epochs $\leq10$ in both $g$ and $r$ bands.
The bronze class is defined as having $g$- and $r$-band LCs consisting of data points  $\leq10$\,epochs.
The gold class is defined by the following two criteria: {\bf [1]} no gap longer than 20 days in rest frame, but very late-time gaps ({\it i.e.} taken at 100 days after the peak or 2 mag fainter than the peak) are allowed. {\bf [2]} the LC covers phases which reach at least 0.5 and 1.0 mag below the peak luminosity pre- and post-peak, respectively. The gold class has 40 SLSNe-I and the bronze class has only 4. The remaining 33 events are in the silver class whose LCs have $>10$ epochs in either $g$ or $r$ band, but do not meet both gold class criteria. Some analysis is performed only with the LCs in the `gold' and `silver' classes.

\section{Light Curve Modeling} 
\label{sec:modeling}

\subsection{Velocity Measurements at Peak Phases} \label{subsec:velocity}
The width of a bolometric LC is closely related to the effective diffusion time scale, which describes the time photons take to travel through the ejecta material and is proportional to $\left( M_{\rm ej}/V_{\rm ej} \right) ^{1/2}$. When modeling LCs to derive ejecta masses and other physical parameters, it reduces the number of free parameters and uncertainties if $V_{\rm ej}$ can be constrained separately from optical spectra. Here $V_{\rm ej}$ is approximated with the photospheric velocity at peak phases \citep{Arnett_1982,Kasen2010, Nicholl2017}. 

We use three different ways to measure the photospheric velocities. The first method is to use \ion{Fe}{2}\,$\lambda\lambda\,4924, 5018, 5169$ absorption lines as tracers. \citet{Liu2016} and \citet{Modjaz_2016} have shown that this method can derive robust measurements for stripped-envelope SNe and the spectral template-matching technique can mitigate blended \ion{Fe}{2} lines for high-velocity events such as SNe\,Ic-BL. The second type of spectral tracers is the five \ion{O}{2} absorption lines at $3737 - 4650$~\AA, the hall-mark features for SLSNe-I at early phases \citep{Quimby2011}. These are useful for velocity measurements as shown by \citet{Quimby_2018,Gal-Yam_2019b}. The third method is to cross-correlate the spectra of our events with spectral templates from several well-studied SLSNe-I, and estimate the relative spectral shifts, thus their relative velocities.

\begin{figure}[htp]
\includegraphics[width=0.5\textwidth]{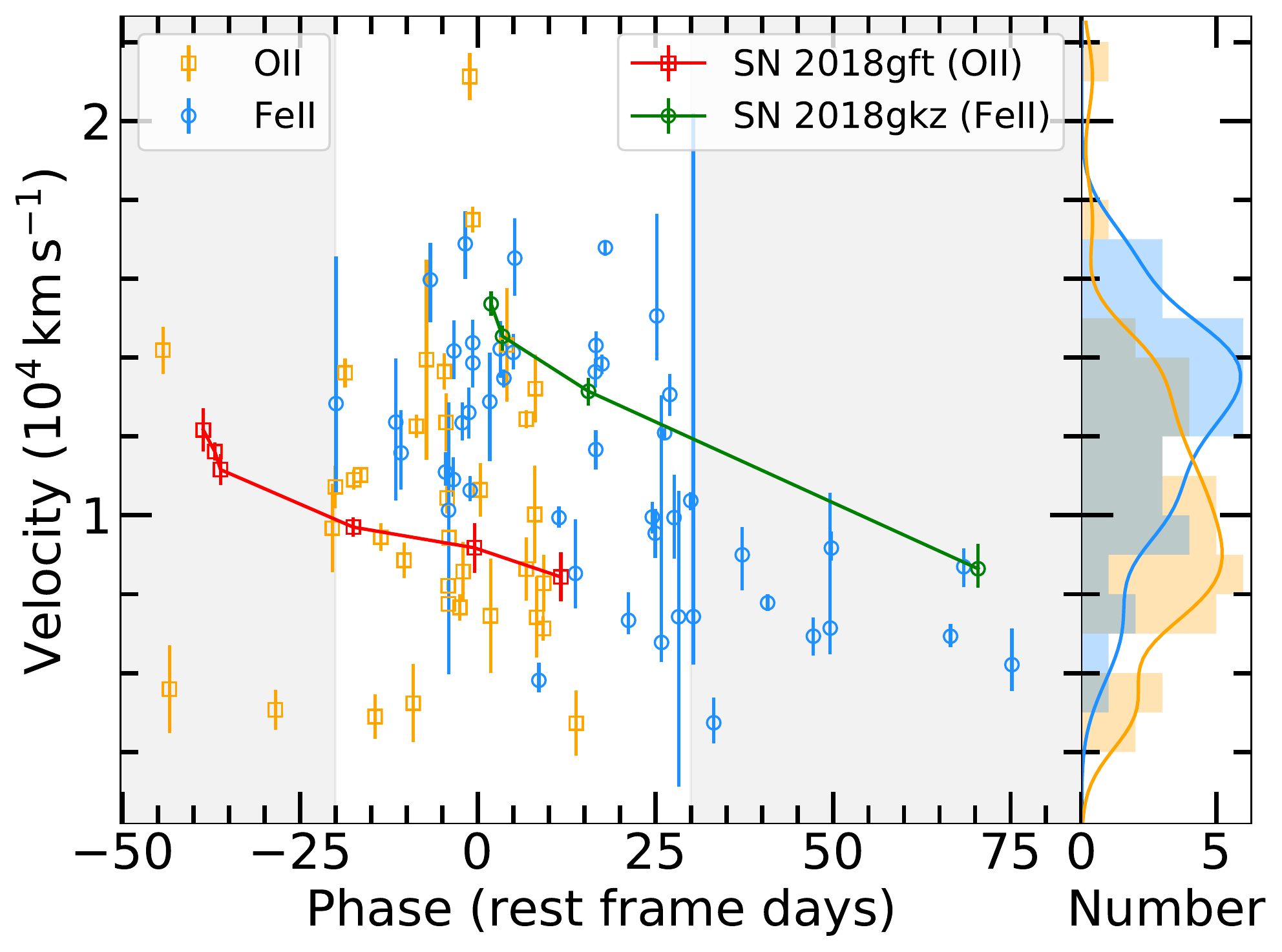}
\caption{Photospheric velocities versus phases. \ion{Fe}{2} and \ion{O}{2} velocities are marked with circles and squares, respectively. We mark the early (${\rm phase} < -20$\,days) and late (${\rm phase} > +30$\,days) time region with shaded area. The velocity evolutions of two events with good phase coverage are highlighted in different colors. The right panel shows the distribution of \ion{Fe}{2} and \ion{O}{2} velocities near peak phases in the white area. The blue and orange solid lines show the kernel density estimation of the distributions, respectively.}
\label{fig:velocity}
\end{figure}

When measuring the velocities, the spectra are first cleaned by removing the narrow host emission lines, smoothed and divided by the continua. 
Using the template-matching code {\tt SESNspectraLib}\footnote{\url{https://github.com/nyusngroup/SESNspectraLib}} \citep{Bianco_2016,Liu2016,Modjaz_2016}, we measure 51 velocities with errors from Markov Chain Monte Carlo (MCMC) for 33 SLSNe-I using \ion{Fe}{2} features. Note that the pre-peak \ion{Fe}{2} velocity can be underestimated due to the contamination of \ion{Fe}{3} as illustrated in \citet{Liu2017}. And the post-peak \ion{Fe}{2} velocity can be overestimated by about $9000 \,{\rm km\, s^{-1}}$ due to the blending effect of \ion{Fe}{2}\,$\lambda\lambda\,4924, 5018$ and \ion{Fe}{2}\,$\lambda\,5169$ \citep{Quimby_2018}. We excluded those ambiguous measurements with velocities $> 15,000 \,{\rm km\, s^{-1}}$ and only one broad absorption component near the Fe II absorption features. When measuring \ion{O}{2}, we derive the spectral shifts by fitting the local minima using the least-squares method, assuming the five absorption features have the same velocity. The errors are calculated from the co-variance matrix. The five \ion{O}{2} features do not have the same strength, with \ion{O}{2} $\lambda\lambda\,4358,4651$ (features A \&\ B) the strongest. For some spectra, we fit only 2 to 4 significant features since the others are too weak. We derive 41 velocities at phases $-43$ to $+14$\,days for 33 SLSNe-I using \ion{O}{2} tracers. SN\,2018gft has very strong \ion{O}{2} features from $-38.6$ to $+11.7$ days and we highlight its evolution in Figure~\ref{fig:velocity}.

For six events (namely SN\,2018kyt, SN\,2019ujb, SN\,2019zbv, SN\,2019aamr, SN\,2019aamt and SN\,2020afag) in our sample, their \ion{O}{2} lines are not clearly identified, especially when \ion{Mg}{1} $\lambda\lambda\,3829,3832,3838$ or \ion{Ca}{2} $\lambda\lambda\,3934,3969$ may be present. For these spectra with ambiguous \ion{O}{2} feature identifications, we match them with three well-observed SLSNe-I, PTF12dam, SN\,2011ke and SN\,2015bn near peak phases \citep{Inserra_2013,Quimby_2018,Nicholl2013,Nicholl_2016b}. We record the velocities derived from the five best-matching templates and use their mean value as our final result and the standard deviation as the error. 

In total, we are able to measure photospheric velocities near peak phases for 51 events. The remaining events do not show clear \ion{Fe}{2} or \ion{O}{2} features or do not have sufficient spectra at the right phases. The measured velocities $V_{\rm ph}$ are listed in Table~\ref{tab:v}.

Figure~\ref{fig:velocity} shows the measured velocities as a function of phase as well as the histograms of velocity distributions. The shaded region marks the early (${\rm phase} < -20$\,days) and late (${\rm phase} > +30$\,days) time region. The histogram distributions in the right panel show that near the LC peak ($-20\,{\rm days} \lesssim {\rm phase} \lesssim +30$\,days), the \ion{Fe}{2} velocity has a median value of 12,800$ \,{\rm km\, s^{-1}}$, whereas the median \ion{O}{2} velocity is only 9700$ \,{\rm km\, s^{-1}}$. Considering the velocities from both ionic species, the median peak photospheric velocity is about 10,900$ \,{\rm km\, s^{-1}}$ for our sample of SLSNe-I. A similar trend is found for PTF12dam, where the \ion{O}{2} velocity at peak is 3000$ \,{\rm km\, s^{-1}}$ slower than that of \ion{Fe}{2} \citep{Quimby_2018}. This is an illustration that Fe$^+$ ions need lower ionization temperatures and are located at the outer layers of ejecta, thus having higher velocities, whereas the O$^+$ ions tend to be in the inner ejecta regions with lower velocities. To avoid possible biases caused by the choice of binning for the histogram distribution, we apply kernel density estimation on all the histograms in this paper (shown as the solid lines in the histograms) using a Gaussian kernel offered by the machine learning package Scikit-learn \citep{scikit-learn}.

\subsection{Light Curve Modeling} \label{subsec:mosfit}
\subsubsection{Model setup}
\label{subsubsec:2model}
One of the primary science goals in this paper is to set constraints on the power sources for the luminous optical emission seen in the SLSNe-I. The open source software {\tt MOSFiT} \citep{Guillochon2018} is used to model the LCs of 70 (out of 77) events in the gold and silver class with good phase coverage. We exclude 3 silver and all 4 bronze events with poorly sampled data before the peak. For another 7 events, we exclude the faint data obtained at either very early or late phases. 
%SN\,2020htd, one of the 8 events, drops below 21 mag (measured from Liverpool telescope) at around $+75$\,days post peak and shows a clear bump after it. The data after $+75$\,days are reaching the detection limit of ZTF and thus excluded in this paper.
The excluded regions for 7 events are listed in Table~\ref{tab:mosfitrange}. For all the other events, we include all the data in the fit.  
%which apparently cannot be fit by a single model together with their peak.
%Two special events, {\it i.e.} SN\,2020fvm and SN\,2019stc, show a strong secondary peak. We limit the range of LC modeling to their first peak only. The excluded regions are listed in Table~\ref{tab:mosfitrange}. 
The input LC data are corrected for Galactic extinction, but not the host extinction, which is a free parameter in {\tt MOSFiT}. 
ZTF sometimes can have multiple detections in the same band on one night. The input LC data are binned into one-day bin, to reduce the number of input data in the MCMC fit and avoid over-weighting the epochs with multiple detections. Those data points that deviate obviously from those on the same epoch will be regarded as outliers instead of real undulations in this paper.
%The input LC data are binned into one-day bins to reduce the complexity of the fit and avoid the effect from occasional deviation points of the observations.

\begin{center}
\begin{longtable}[htp]{lc|lc}
%\begin{table}[htp]
\caption{Excluded regions in LC modeling.}\\
%\begin{tabular}{lc|lc}
\toprule
\multirow{2}{*}[-4pt]{Name} & MJD & \multirow{2}{*}[-4pt]{Name} & MJD \\
& (days) & & (days) \\
\midrule
SN\,2018hpq & $<58385$ & SN\,2020onb & $<59030$ \\ 
SN\,2018lzv & $<58275$ & SN\,2020xkv & $<59070$ \\ 
SN\,2019eot & $<58595$ & SN\,2019szu & $>59090$ \\ 
SN\,2020exj & $<58920$ &  & \\
\bottomrule
%\end{tabular}
%\hline
%\bottomrule
\label{tab:mosfitrange}
%\end{table}
\end{longtable}
\end{center}

We run the {\tt MOSFiT} via Dynamic Nested Sampling \citep[{\tt dynesty},][]{Speagle_2020} %and request that both the initial $\Delta$log$_{10}Z$ in the static phase and the KL-divergence criterion in the dynamic phase equal 0.02. %until each run has converged. The convergence is controlled by the {\tt --run-until-converged} option, and we set it to the default value 0.02 (see their documents\footnote{\url{https://mosfit.readthedocs.io/en/latest/fitting.html#nested-sampling}} for details). 
until each run has converged under the default stopping criterion (see their documents\footnote{\url{https://mosfit.readthedocs.io/en/latest/fitting.html##nested-sampling}} for details).
We choose two commonly used models, {\it i.e.} the magnetar \citep[{\it slsn} model in {\tt MOSFiT},][]{Kasen2010,Nicholl2017b} and the CSM+Ni \citep[{\it csmni} model in {\tt MOSFiT},][]{Chatzopoulos2013,Villar2017,Jiang2020}. For the CSM+Ni model, we fit both a constant density ($s=0$) and a wind-like density ($s=2$) profile. The key parameters are listed in Table~\ref{tab:modeling}. Each free parameter has a prior distribution defined by {\tt MOSFiT}. These prior distributions can be modified for specific dataset. For example, for most of our sources, the velocities are measured from the spectra, and we set the $V_{\rm ej}$ prior to a flat distribution from %$0.7 - 1.3$ times the measured peak velocities. For some events with only post-peak velocities, we increase this multiplication factor to $1.6$ when setting the upper limit of the $V_{\rm ej}$ prior.
$0.6 - 1.6$ times the measured velocities at the peak. 
Following the assumption of \citet{Nicholl2017b}, we use the \ion{Fe}{2} velocity to set the prior of $V_{\rm ej}$ . Considering the systematic difference between \ion{Fe}{2} and \ion{O}{2} velocities (shown by \citet{Quimby_2018} and our measurements in \S~\ref{subsec:velocity}), we add a correction of $3000 \,{\rm km\, s^{-1}}$ when using \ion{O}{2} velocity. 
By testing the \ion{Fe}{2} and \ion{O}{2} velocities in \citet{Quimby_2018} and \citet{Liu2017}, we find this range is wide enough to counteract the influence of underestimating pre-peak \ion{Fe}{2} velocities and  the velocity evolution due to the phase differences between the spectra and LC peaks. %the systematic velocity difference between \ion{Fe}{2} and \ion{O}{2} lines. 
For the events without measured velocities, we use a constant velocity range of $3000-25000 \,{\rm km\, s^{-1}}$, allowing {\tt MOSFiT} to estimate velocities from the LC fitting.

For the magnetar model, we set the angle between the magnetic field and the spin axis $\theta_{\rm PB}=\pi/2$. The output $B$-field from the {\tt MOSFiT} is only the perpendicular component $B_{\rm \perp}$, which relates to the total magnetic field through $B_{\rm total} = B_{\rm \perp}/\sin\theta$. For the priors of the other parameters in the magnetar model, we use probability distributions similar to the ones in \citet{Nicholl2017b}. For the CSM+Ni model, we use the default distributions of CSM mass and CSM shell density. We set the mass ratio of $^{56}$Ni to be less than $50\%$, the radius of the progenitor star to have a range from 0.01 to 100 AU and the opacity $\kappa$ from 0.05 to 0.34 $\rm g\,cm^{-2}$. Except for $\kappa$, the other parameters common to both the magnetar and CSM+Ni models have the same prior distributions for consistency. Another parameter -- $\gamma$-ray photon leakage parameter $\kappa_{\gamma}$ -- has a constant prior of ${\rm log_{10}}\kappa_{\gamma}$ between (-2, +2), as used in \citet{Nicholl2017}. %We do not find necessary to use a wider range of the prior for $\kappa_{\gamma}$, as we discuss the $\kappa_{\gamma}$ distribution for the full sample below.

Finally, for each run, {\tt MOSFiT} outputs a large number of possible model LCs with different weights based on {\tt dynesty}. We use the weighted median LC as the final model LC and evaluate the error by 16\% and 84\% percentiles. In fitting the observed LCs of SN\,2018don with the CSM+Ni model, {\tt MOSFiT} converges to different outputs with the same priors for different runs. This is primarily due to the large parameter space where the convergence could be at local minimum. In this case, we ran {\tt MOSFiT} several times and use the best result indicated by the smallest reduced $\chi^2$ parameter as defined below.

\subsubsection{Importance of ejecta-CSM interaction in SLSNe-I}
\label{subsec:csmlc}

One basic question is which of these two models fits the data better. To quantify this, we use the reduced $\chi^2$ parameter using the numbers of fitted parameters -- 11 and 12 for the magnetar and the CSM+Ni model respectively. 
%A small number of data points from P200 and LT have very high signal-to-noise ratio (SNR) and small magnitude errors. To prevent excessive weights on these data, we set the smallest error to 0.01\,mag when calculating the $\chi^2$ parameter.

\begin{figure}[htp]
\includegraphics[width=0.5\textwidth]{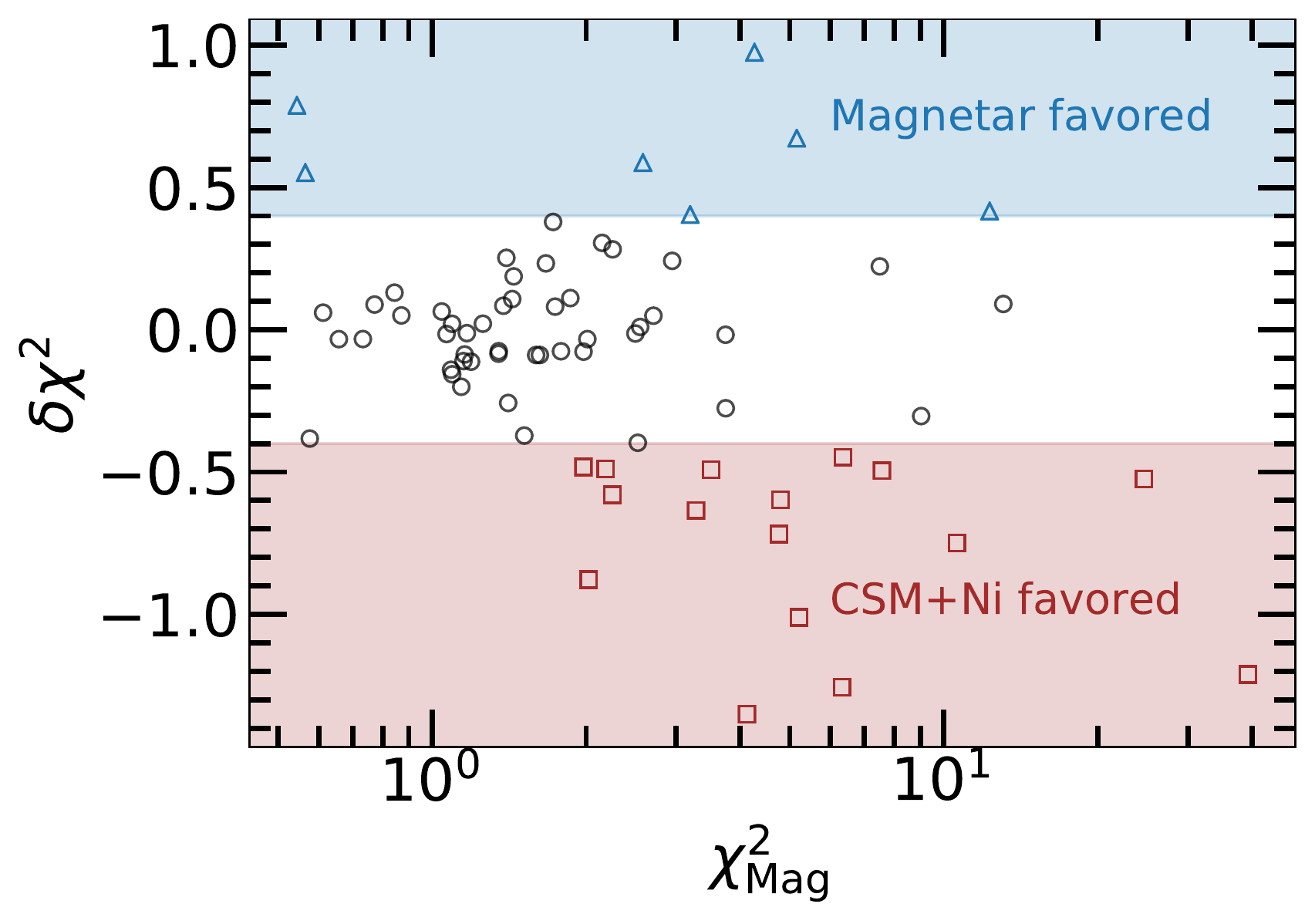}
\caption{Reduced $\chi^2$ values of the magnetar and CSM+Ni modeling. Y-axis shows the relative differences of $\chi^2$ between these two models. The red area marks the region where the events are better fit by the CSM+Ni model, {\it i.e.} $\delta \chi^2 < -40\%$, while the blue area marks the magnetar-favored region, {\it i.e.} $\delta \chi^2 > 40\%$. The events in different areas are labeled with different symbols.} 
\label{fig:chi}
\end{figure}

Figure~\ref{fig:chi} shows the $\chi^2_{\rm Mag}$ computed for the magnetar model versus the $\chi^2$ difference between the CSM+Ni model and the magnetar model. For the CSM+Ni model, we choose either $s=0$ or $s=2$ depending on which $\chi^2$ is smaller. The large absolute $\chi^2$ values can be due to underestimated photometric errors and the LC undulations. 
%For example, the strongly undulating event SN\,2019unb, located at the very right of Figure~,  can not be properly fit by either model.  
According to $\delta \chi^2 = (\chi^2_{\rm CSM} - \chi^2_{\rm Mag})/[(\chi^2_{\rm CSM} + \chi^2_{\rm Mag})/2]$, we find that only a small fraction of the 70 SLSN-I events clearly prefer one model, with 16 events better fit by the CSM+Ni model and 7 by the magnetar model. The majority ($47/70=67\%$) of the sample can be equally well fit by both models ($|\delta \chi^2| < 40\%$). This indicates that LCs alone can not unambiguously distinguish between these two energy sources.  

\begin{figure*}[t!]
\includegraphics[width=\textwidth]{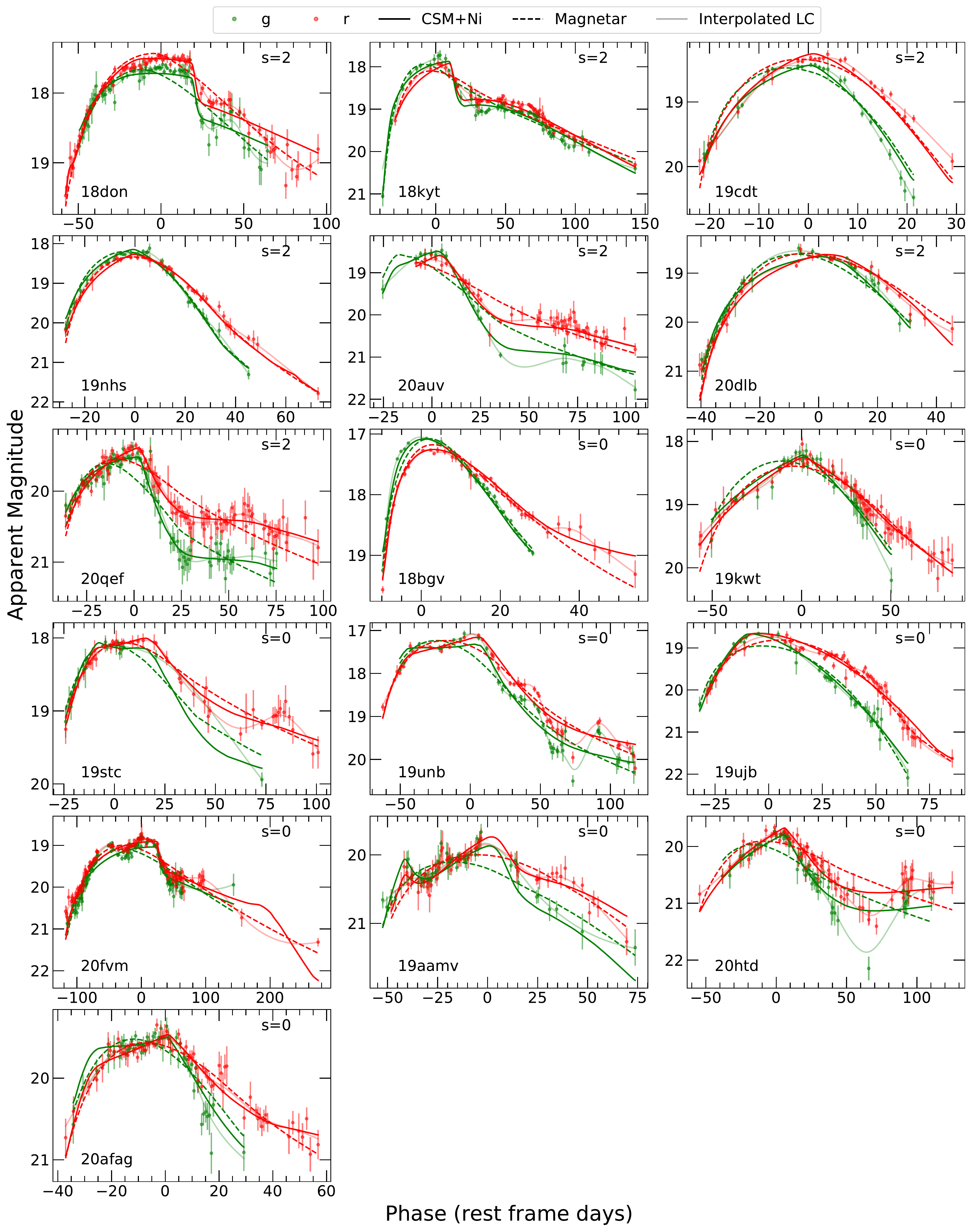}
\caption{The 16 events best fit by the CSM+Ni model. CSM+Ni model, magnetar model and GP interpolated LCs are plotted with solid lines, dashed lines and translucent lines, respectively. The type of CSM structure, constant density $(s=0)$ or wind-like $(s=2)$, is labeled at the top right corner. The data in shaded region are masked out during the LC modeling.}
\label{fig:CSM}
\end{figure*}

The 16 events favoring the CSM+Ni model have several distinct features. First, some LCs show a steep flux drop after the primary peak, {\it e.g.} SN\,2018don and SN\,2020afag.  This steep change of LC slope generally can not be reproduced by the magnetar model, as revealed by the poor fits (see the dashed lines) in Figure~\ref{fig:CSM}. However, the CSM+Ni model with a wind-like ($s=2$) or constant ($s=0$) density profile does much better (solid lines in Figure~\ref{fig:CSM}). The rapid decline has a simple physical explanation where the forward shock has run through the CSM \citep{Chatzopoulos2012}. Second, some LCs have inverted V-shaped evolution, {\it i.e.} linear rise and decline with a sharp peak, {\it e.g.} SN\,2019kwt and SN\,2020htd. This type of LCs can also be better fit by the CSI model, with a constant density CSM shell ($s=0$), as previously noted by \citet{Chatzopoulos2013} and \citet{Wheeler2017}. 
%This is consistent with the finding shown in Paper I that most events favoring the CSM+Ni model tend to have longer rise times but shorter decay times compared with the other SLSNe-I.

\begin{figure}[htp]
\includegraphics[width=0.5\textwidth]{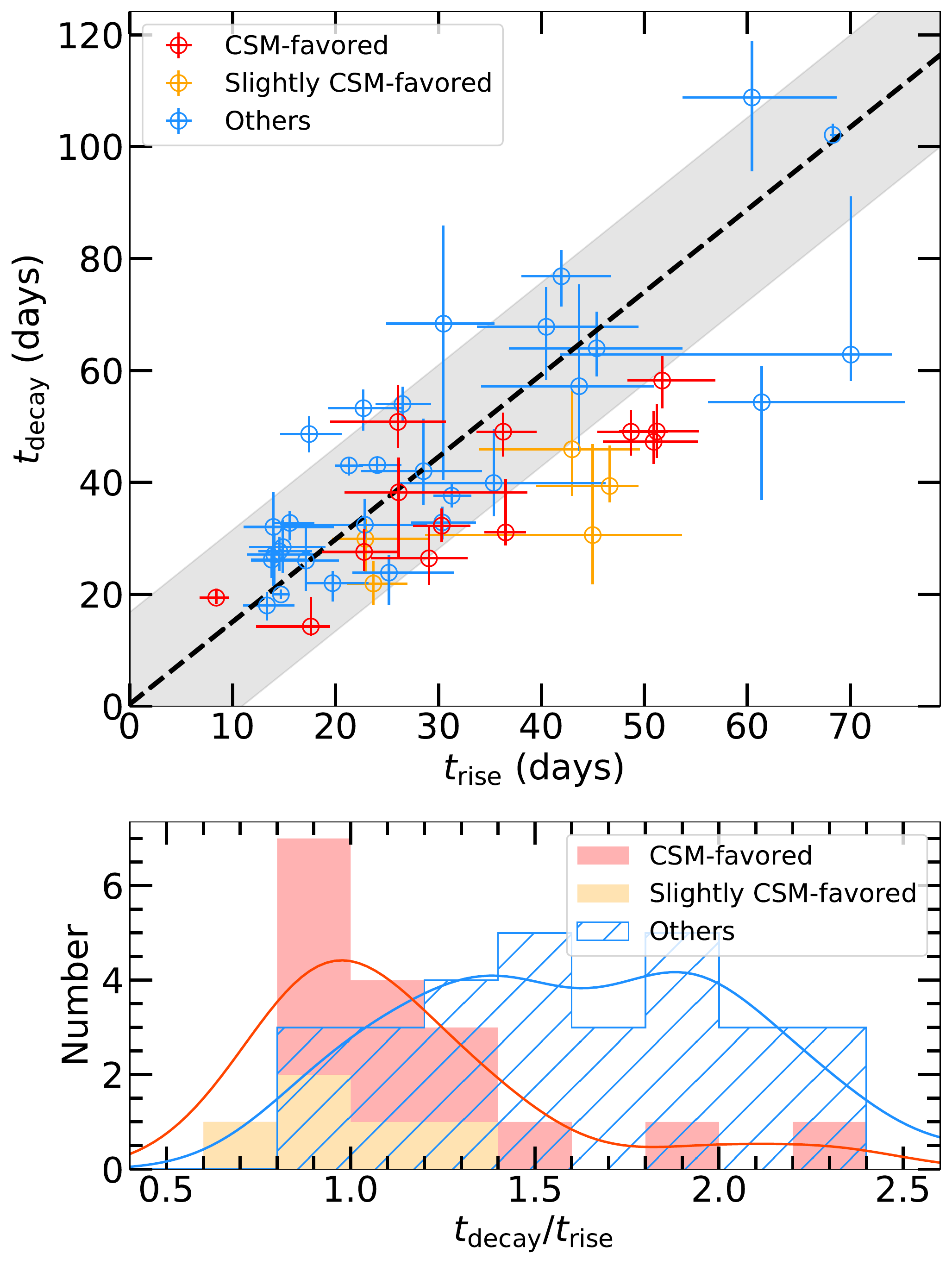}
\caption{Time scales of ZTF sample (from the Figure~5 in Paper I). In the upper panel, x/y axes are the rise/decay time scales. The red, yellow, blue dots represent the CSM-favored events ($\delta \chi^2 < -40\%$), the ones slightly favored by the CSM+Ni model ($-40\% \leq \delta \chi^2 < -20\%$) and the others, respectively. The dashed line shows the linear fit between rise and decay time scales and the corresponding $1\sigma$ error. The lower panel shows the distribution of the $t_{\rm decay} / t_{\rm rise}$ ratio. The orange solid line shows the kernel density estimation of the two CSM-favored groups, and the blue line shows that of the others.}
\label{fig:decline2rise2}
\end{figure}

The steep flux drop can also be characterized by the time scales of SLSNe-I. In Paper I, we measured rise and decay time scales of the LCs for our sample and confirmed that the rise and decay time scales roughly follow a linear relation, {\it i.e.} $t_{\rm decay}=1.47~t_{\rm rise}+0.35$\,days. 
%While most events follow this relation, some events lie below the linear relation, {\it i.e.} having either longer rise time or shorter decay time. 
In Figure~\ref{fig:decline2rise2}, we plot the rise and decay time scales of our sample (similar to Figure~5 in Paper I) and highlight the SLSNe-I favored by the CSM+Ni model. Most CSM-favored events are below the linear relation. They tend to have longer rise times and shorter decay times compared with those favored by the magnetar model or equally well fit by both models. 
Such a trend becomes more significant when including those that can be properly fit by both the CSM+Ni and magnetar models but with the former scenario being slightly favored ({\it i.e.} $-40\% \leq \delta \chi^2 < -20\%$). 
We applied a two-sample Kolmogorov-Smirnov test on the ratio of rise and decay times between CSM-favored events and the others. The result ($D = 0.54, p = 0.001$) shows that the CSM-favored events and the others indeed have different distributions. So we conclude that the CSM-favored SLSNe-I tend to have longer rise time and faster decay time.

We infer that the {\bf minimum} fraction favoring the CSM+Ni model in our sample is $23\%$ (16/70). Half (8) of these events have smooth LCs, but clearly prefer the CSM+Ni model. Our analysis in \S\ref{subsec:LCshape} shows that CSI likely plays an important if not dominant role in all sources with LC undulations. In \S\ref{subsec:LCshape}, we quantitatively identify 17 events from our sample have either weak or strong undulations. If taking into account all 17 undulating sources plus the 8 events with smooth LCs and favoring the CSM+Ni model, the fraction of CSM powered events can be as high as $25-44\%$ (25/73) at a confidence level (CL) of 95\%\ \citep{Gehrels_1986}. Such a high fraction implies that H-poor (some also He-poor) CSM around SLSNe-I and CSI are quite common. %and that CSI is much more prevalent than previously thought.

\subsubsection{Physical Parameters Derived from Model Fittings}
\label{subsec:parameters}

We compare the peak luminosities and temperatures derived from {\tt MOSFiT} with that from the SED fitting in Paper I, and find that they are largely consistent with each other, with small offsets of $2^{+18}_{-11}\%$\, and $-1^{+9}_{-12}\%$ respectively. Figure~\ref{fig:mosfitcompare} displays several relations between the derived parameters, similar to \citet[][their Figure 6]{Nicholl2017b}. $E_{\rm k}$ is derived from $M_{\rm ej}$ and $V_{\rm ph}$, assuming $E_{\rm k}=0.3 M_{\rm ej} V_{\rm ph}^2$. This relation is valid for a homogeneous density profile, and also adopted for the CSM+Ni model. Our $E_{\rm k}$ values are thus somewhat lower than those derived by \citet{Nicholl2017b} which used $E_{\rm k}=1/2 M_{\rm ej} V_{\rm ph}^2$. The overall distributions of $P, B_{\rm \perp}, M_{\rm ej}$, and $E_{\rm k}$ are similar to the results in \citet{Nicholl2017b, Blanchard2020, Hsu2021}. 
We test the correlations between these four parameters using the Spearman rank correlation coefficient. 
As also found by \citet{Blanchard2020} and \citet{Hsu2021}, $M_{\rm ej}$ shows a strong negative correlation ($\rho=-0.53,~ p<10^{-5}$ in our sample) with magnetar spin period $P$, indicating that SLSNe-I with smaller ejecta masses require less central power with slower spinning neutron stars. 

\begin{figure}[htp]
\centering
\includegraphics[width=0.5\textwidth]{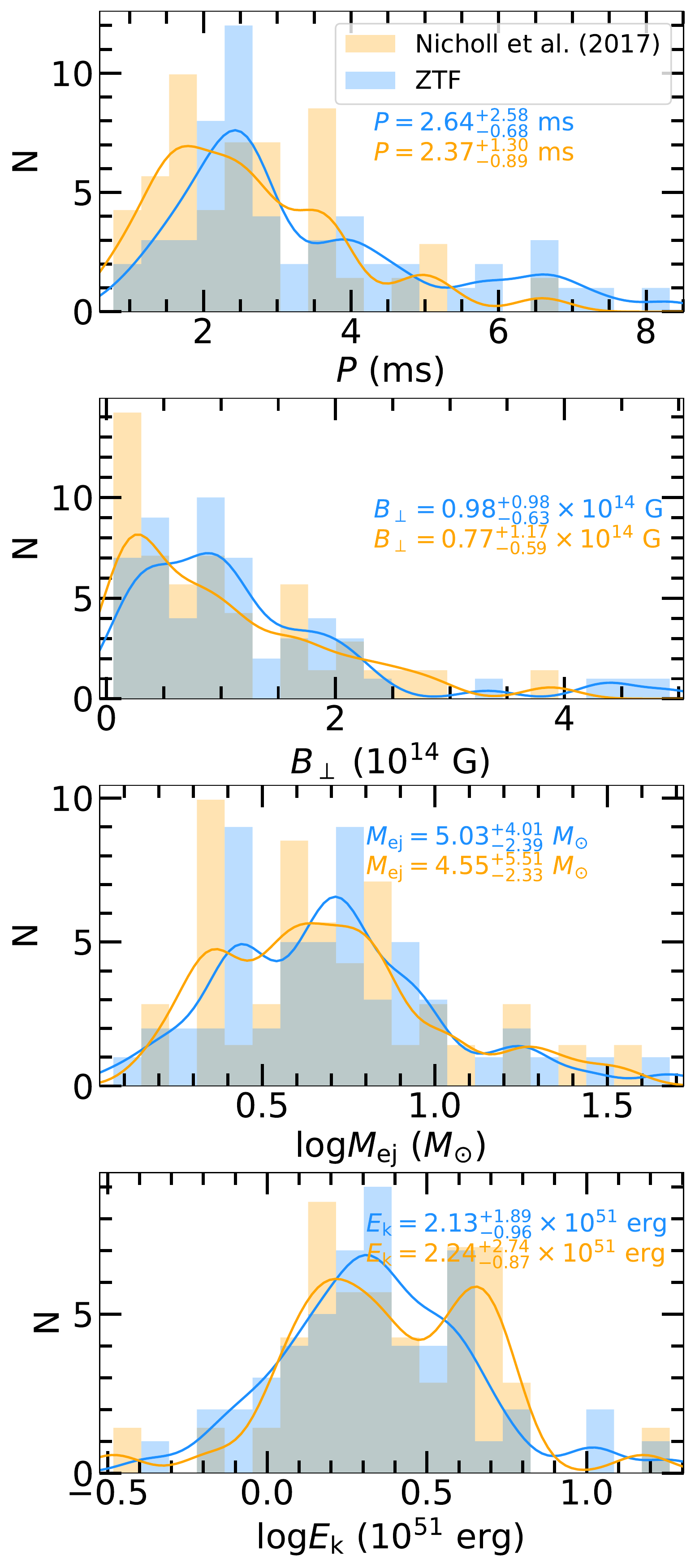}
\caption{Distribution of a series of important parameters ($P, B_{\rm \perp}, M_{\rm ej}, E_{\rm k}$) in the magnetar model. The blue color indicates the distribution and median value for the 54 (7+47) SLSNe-I fit by the magnetar models (7 events favor the magnetar models, and 47 can be fit equally well by both the CSM+Ni and magnetar models). Those from \citet{Nicholl2017b} are shown in orange for comparison (total number is normalized to that of ZTF sample). The solid lines show the kernel density estimation of the distributions.}
\label{fig:mosfitdistribution}
\end{figure}

\begin{figure}[htp]
\centering
\includegraphics[width=0.5\textwidth]{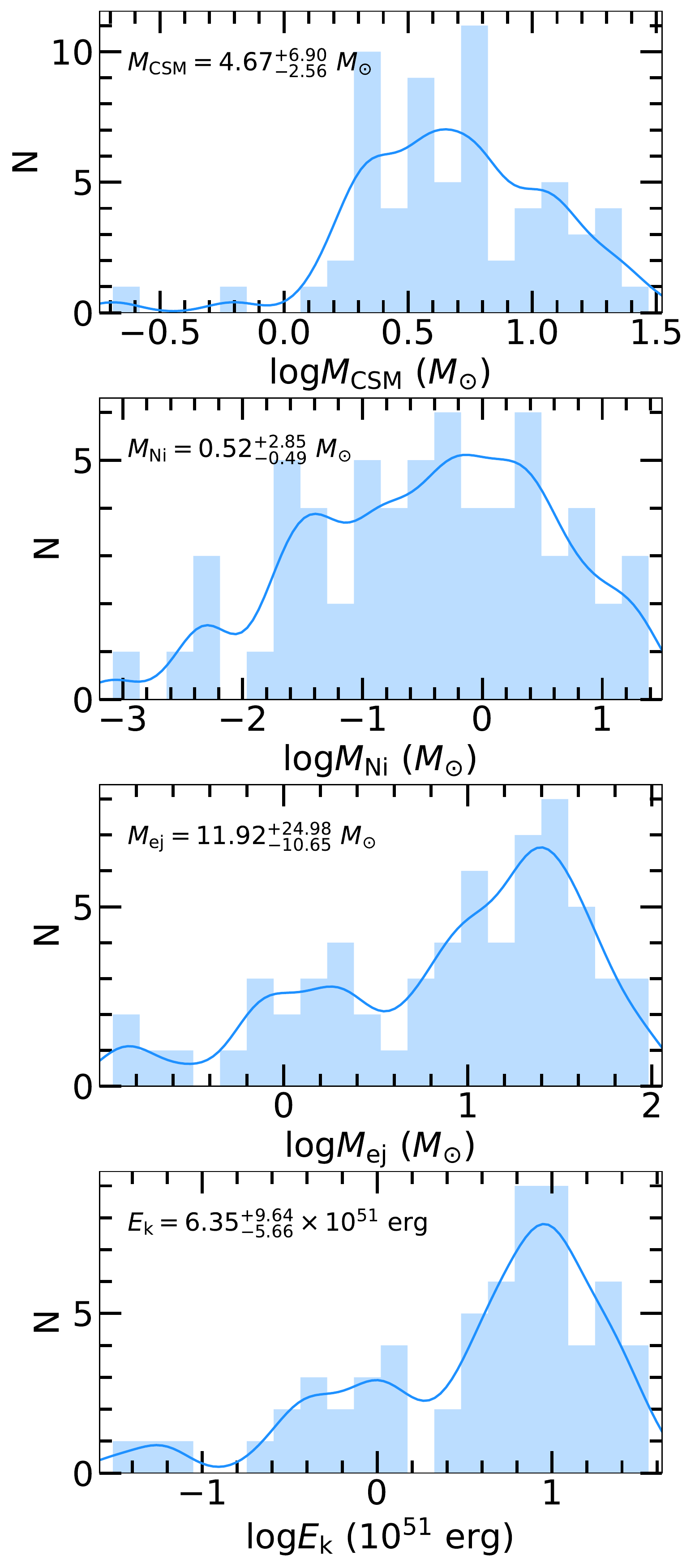}
\caption{Distribution of key parameters ($M_{\rm ej}, M_{\rm CSM}, M_{\rm Ni}, E_{\rm k}$) derived for the CSM+Ni model. Here we include the 16 SLSNe-I favored by the CSM+Ni models and the 47 events which can be fit equally well by both the CSM+Ni and magnetar models. The solid lines show the kernel density estimation of the distributions.}
\label{fig:CSMmosfitdistribution}
\end{figure}

Figures~\ref{fig:mosfitdistribution} and \ref{fig:CSMmosfitdistribution} show the distributions of the key parameters from the magnetar model ($P, B_{\rm \perp}, M_{\rm ej}$, and $E_{\rm k}$) and the CSM+Ni model ($M_{\rm ej}, M_{\rm CSM}, M_{\rm Ni}$, and $E_{\rm k}$). The median values and the 1$\sigma$ errors (16\% and 84\% percentiles) of the key parameters from the two models are listed in Table~\ref{tab:modelparameter}. The CSM+Ni model we use in {\tt MOSFiT} is based on the semi-analytic model from \citet{Chatzopoulos2012,Chatzopoulos2013}. However, it has been shown that the semi-analytic model and hydrodynamic simulations can produce inconsistent results assuming the same CSM structure \citep{Moriya2013,Sorokina2016}. The quantitative value of the CSM parameters from the CSM+Ni model could be just an order of magnitude estimate.

\begin{center}
\begin{table}[htp]
\begin{longtable}[b!]{lcc}
%\begin{table}[htp]
\caption{Medians of key modeling parameters}\\
%\begin{tabular}{p{3cm}<{\centering}p{2cm}<{\centering}p{2cm}<{\centering}}
%\begin{tabular}{lcc}
\toprule
Parameter & Magnetar & CSM+Ni \\
\midrule
$P \ (\mathrm{ms})$ & $2.64^{+2.58}_{-0.68}$ & - \\
$B_{\rm \perp} \ (10^{14}\ \mathrm{G})$ & $0.98^{+0.98}_{-0.63}$ & - \\
$M_{\rm CSM} \ (M_{\odot})$ & - & $4.67^{+6.90}_{-2.56}$ \\
$M_{\rm Ni} \ (M_{\odot})$ & - & $0.52^{+2.85}_{-0.49}$ \\
$M_{\rm ej} \ (M_{\odot})$ & $5.03^{+4.01}_{-2.39}$ & $11.92^{+24.98}_{-10.65}$ \\
$E_{\rm k} \ (10^{51}\ \mathrm{erg})$ & $2.13^{+1.89}_{-0.96}$ & $6.35^{+9.64}_{-5.66}$ \\
\bottomrule
%\end{tabular}

\label{tab:modelparameter}
%\end{table}
\end{longtable}
\end{table}
\end{center}

\begin{figure}[htp]
\includegraphics[width=0.5\textwidth]{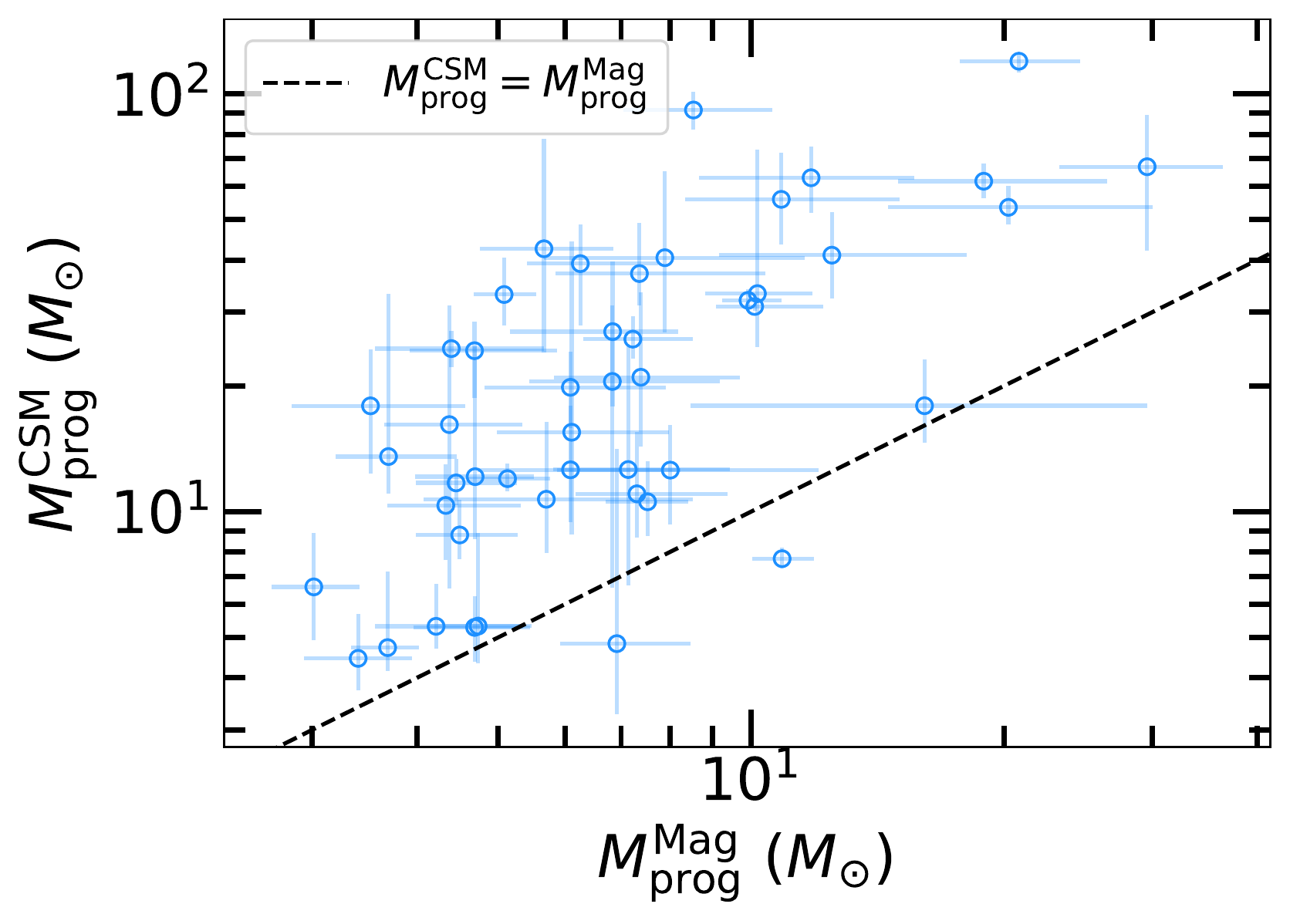}
\caption{Progenitor mass $M_{\rm prog}$ derived from the magnetar model (x-axis) and from the CSM+Ni model (y-axis) for the 47 SLSNe-I for which both models are equally good fits.}
\label{fig:progenitor}
\end{figure}

Compared with the magnetar model, the $M_{\rm ej}$ estimates from the CSM+Ni model are significantly higher. The final mass of the progenitor star $M_{\rm prog}$ is estimated by summing up $M_{\rm ej}$ and neutron-star mass for the magnetar model, and $M_{\rm ej}$, $M_{\rm CSM}$ plus a typical neutron-star mass \citep[$1.4\,M_\odot$,][]{Lattimer2007} for the CSM+Ni model. Note that the progenitor mass calculated with this method is just a lower limit. Figure~\ref{fig:progenitor} shows the mass estimates for the 47 events which are equally well fit by both the magnetar and CSM+Ni models. The progenitor mass derived from the magnetar model has a median value of $6.83^{+4.04}_{-2.45}\ M_{\odot}$ while it is $17.92^{+24.11}_{-9.82}\ M_{\odot}$ from the CSM+Ni model.

Assuming a stellar population with low metallicity (metal fraction $10\%$ solar, {\it i.e.} $1-2 \times 10^{-3}$), we estimate the zero-age-main-sequence (ZAMS) mass using {\it PARSEC} \citep[][]{Bressan_2012}. A ZAMS mass of $18 - 35 \ M_{\odot}$ is needed for the magnetar model while $30 - 130 \ M_{\odot}$ is required for the CSM+Ni model. It is not surprising that the predicted progenitor and ZAMS masses for the CSM+Ni model are much larger. This reflects the fact that the CSM+Ni model needs both larger ejecta and CSM masses in order to supply sufficient energy, as also noted previously by \citet{Chatzopoulos2013}. More importantly, the high ZAMS values for the CSM+Ni model are in the regime ($70 - 140 M_\odot$) where Pulsational electron-positron Pair-Instability Supernova are expected to explode \citep[PPISN,][]{Woosley2017}. PPISN events will also experience violent episodic mass losses. This may naturally explain the presence of substantial amount of CSM. 
On the other hands, around half (9) of the CSM-favored events have kinetic energies higher than $5\times10^{51}$\,erg, which exceed the highest value for PPISN events \citep{Woosley2017}. 
Pair-instability supernova \citep[PISN,][]{Kasen2011,Dessart2012}, which predicts higher kinetic energies and slow-evolving LCs, could explain the three events with rise time longer than 90 days. However, for the rest six events, a central engine \citep[{\it e.g.} magnetar or black hole fallback,][]{Kasen2010, Dexter2013} plus CSI is likely needed to provide such a high energy. 

Another parameter which can significantly impact the ejecta mass estimates is $\kappa_{\gamma}$, the $\gamma$-ray photon leakage parameter. The smaller $\kappa_{\gamma}$ values, the faster LC decays and as compensation, the larger ejecta masses are needed to fit the late-time LCs. The {\tt MOSFiT} derived ${\rm log_{10}}\kappa_{\gamma}$ values for the majority of our sample is $>-1.6$, with a small fraction (8 of 70 for the magnetar model and 2 for the CSM+Ni model) having $-1.96<{\rm log_{10}}\kappa_{\gamma}<-1.6$. 

Recently, \citet{Vurm_2021} carried out three-dimensional Monte Carlo radiative transfer calculations on SLSNe-I using the magnetar model and showed that ${\rm log_{10}}\kappa_{\gamma} = -2$ is an extremely low value for phase $<300$\,days (see their Figure 10). Such low $\kappa_{\gamma}$ requires highly efficient dissipation of the magnetic field or that the spin-down luminosity decays significantly faster than the canonical dipole rate $\propto t^{-2}$ in a way that coincidentally mimics gamma-ray escape. We conclude that our assumed prior (-2, +2) for $\kappa_{\gamma}$ is sufficient and we do not need to explore a wider range of the distribution.

\section{Light Curve Morphologies}
\label{subsec:LCshape}

\subsection{Early double-peak Light Curves} \label{subsec:2peak}
Some SLSNe-I have a weak bump in the early phase, {\it e.g.} SN\,2006oz \citep{Leloudas2012}, LSQ14bdq \citep{Nicholl2015}, DES14X3taz \citep{Smith2016} and SN\,2018bsz \citep{Anderson2018}. \citet{Nicholl2016} speculated that most SLSNe-I may have such early bump features. This was shown to be incorrect by the SLSN-I sample from the Dark Energy Survey (DES) which has very deep photometric limits \citep{Angus2019}. Of the 12 DES SLSNe-I with pre-peak LCs, only 4 showed such a precursor bump.

\begin{figure*}[htp]
\centering
\includegraphics[width=\textwidth]{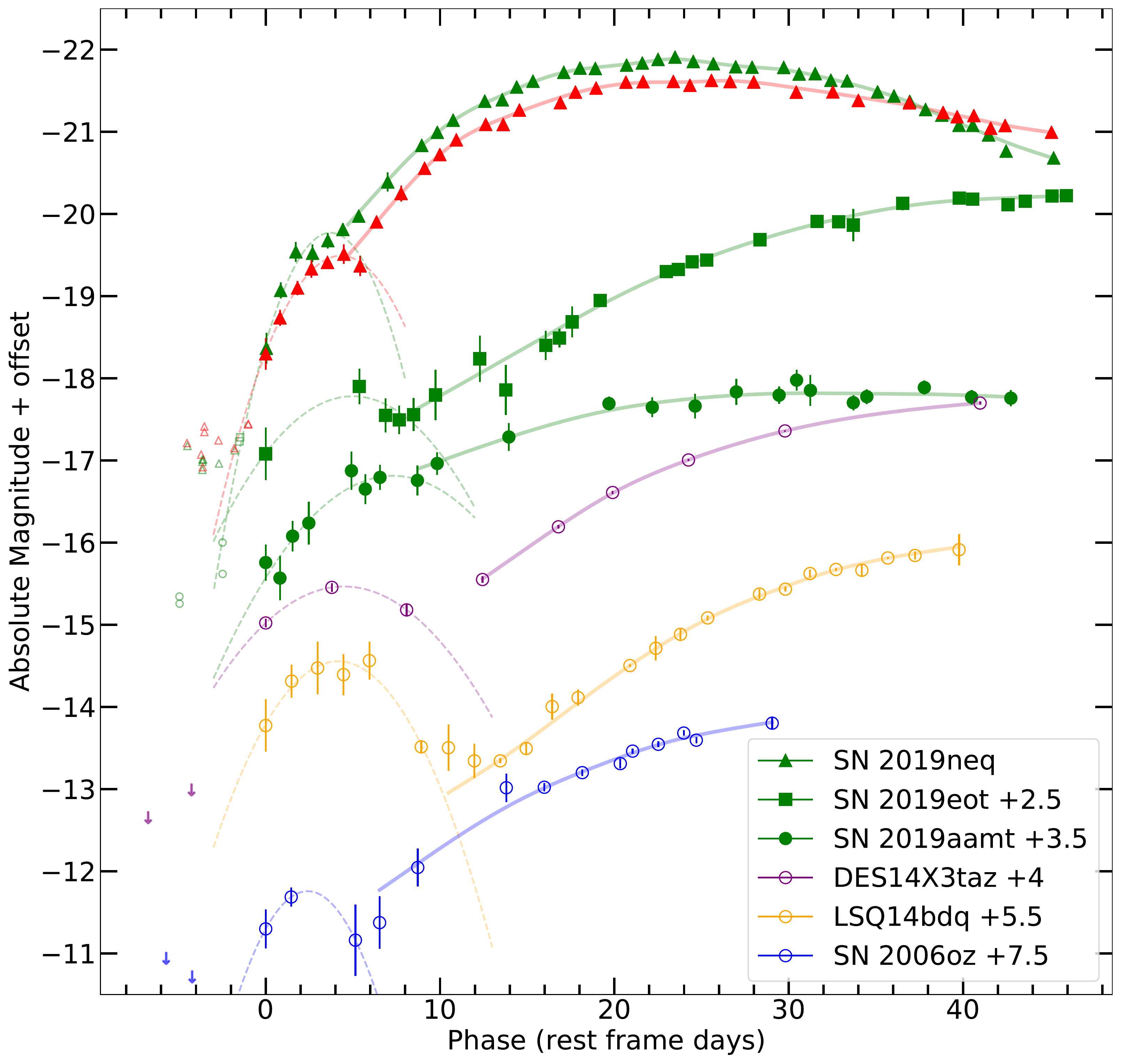}
\caption{Early bump features of three ZTF SLSNe-I and archival SLSNe-I in rest-frame $g$ band. The 3$\sigma$ detections of ZTF events are presented in green solid points while the upper limits are in hollow points with the same markers. SN\,2019neq shows early bumps in both rest-frame $g$ and $r$ bands, and we highlight its $r$-band LC in red. Solid lines show the GP model fits of the LCs. Dashed lines show the second-order polynomial fitting applied to the early bump features.}
\label{fig:double}
\end{figure*}

We also search for early bump features in our SLSN-I sample. We bin the LC data into one-day bins and include only data with $\rm SNR > 3$ in our analysis. We focus on the 15 events which have very early and deep observations after the explosion. These 15 events have at least 4 epochs of pre-peak data which are $1.5$ magnitudes fainter than their main peaks. We find that only three events, SN\,2019eot, SN\,2019neq and SN\,2019aamt, show reliable early-peak features. This corresponds to a fraction of  $6-44\%$ (3/15, CL$=95\%$). This is consistent with the the observed fraction ({\it i.e.} 4/12) in \citet{Angus2019}. The LCs of these three SLSNe-I are shown in Figure~\ref{fig:double}, together with the rest-frame $g$-band LCs of SN\,2006oz, LSQ14bdq and DES14X3taz. The early bumps are detected only in rest-frame $g$ band (observed $r$ band) in SN\,2019eot and SN\,2019aamt,  whereas in SN\,2019neq the initial peak is present in both the $g$- and $r$-band LCs (rest-frame). The early bump is fit by a second-order polynomial whereas the primary peak is fit by the GP method. We define the width of the first peak as the time interval between the two phases when the LC is 2 magnitudes fainter than the peak. The measured time widths and the absolute magnitudes of the first peak are listed in Table~\ref{tab:2peakwidth}. The widths of the early bumps are comparable to the predictions of $10 - 20$\,days from the shock cooling models by \citet{Piro2015}.

\begin{center}
\begin{longtable}[htbp!]{lccc}
%\begin{table}[htp]
\caption{The properties of early bump features}\\
%\begin{tabular}{p{3cm}p{1cm}<{\centering}p{1.5cm}<{\centering}p{1.5cm}<{\centering}}
%\begin{tabular}{lccc}
\toprule
%\multirow{2}{*}{Name} & \multirow{2}{*}{filter$^a$} & Width & \multirow{2}{*}{Name} & Width \\
%& & (day) & & (day) \\
\multirow{2}{*}{Name} & \multirow{2}{*}{Filter$^a$} & Width & $M_{\rm bump}$\\
& & (days$^a$) & (mag) \\
	
\midrule
SN\,2019eot & $g$ & 17.00 & -20.28 \\
SN\,2019aamt & $g$ & 18.56 & -20.31 \\
SN\,2019neq & $g$ & 9.07 & -19.77 \\
SN\,2019neq & $r$ & 11.22 & -19.49 \\
SN\,2006oz & $g$ & 10.04 & -19.26 \\
LSQ14bdq & $g$ & 13.42 & -20.05 \\
DES14X3taz & $g$ & 19.06 & -19.46 \\
\bottomrule
\hspace*{\fill} \\
$^a$Rest frame. \\
%\end{tabular}
\label{tab:2peakwidth}

%\end{table}
\end{longtable}
\end{center}

An alternative model -- magnetar shock breakout through pre-explosion ejecta \citep{Kasen2016} -- can also explain the early bumps of SLSNe-I. In these models, the early bumps are more obvious if the power engine for the primary peak is inefficiently thermalized at the first 15 -- 43\,days after the explosion \citep{Kasen2016,Liu_2021}. The early bumps in SN\,2019neq and SN\,2019aamt are shallower, which perhaps implies that their magnetar energy thermalization is relatively more efficient.

It is worth noting that among the double-peaked SLSNe-I, SN\,2019neq is peculiar, and has a narrow early-bump width (9.07 days) and a short main peak rise time ($\lesssim 25$\,days). This makes it the fastest-evolving SLSNe-I with early bumps up to date. So far most SLSNe-I with early bumps are slow events \citep[rise time $\sim 33 - 100$\,days,][]{Inserra2019}. This could be due to the observational selection bias since fast-evolving events with narrow early bumps can be easily missed by supernova surveys unless with high cadence and early sensitive detections.

\subsection{Undulations in the light curves} 
\label{subsec:Undulation}

\begin{figure}[htp]
\includegraphics[width=0.5\textwidth]{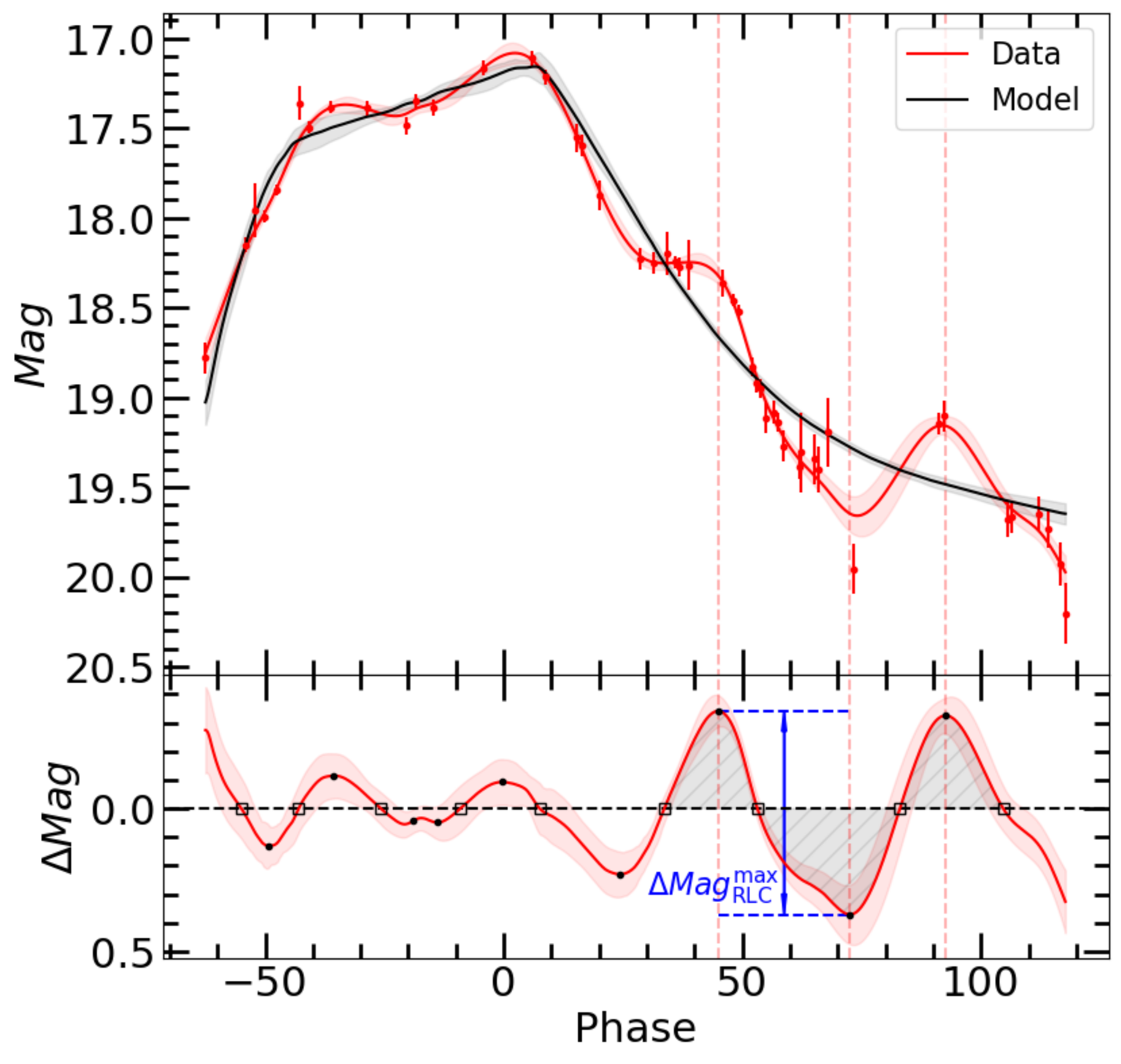}
\caption{An example of the LC undulation measurements. The upper panel shows the $r$-band LC of SN\,2019unb. The GP interpolated LC is plotted with a red line while the black line represents the LC fit with the CSM+Ni (s=0) model. The pink and grey shaded areas represent their 1-sigma errors, respectively. The lower panel shows the residual LC (RLC) with 1$\sigma$ errors from the GP and the magnetar model. The black points represent the local extrema and the blue vertical bar shows $\Delta$Mag$_{\rm RLC}^{\rm max}$, the largest magnitude change between two adjacent extrema of the RLC. The black squares mark the zero points of the RLC. The time duration and energy of the undulations are measured from the intervals between two adjacent zero points as shown in the hatched area. The peaks/dips of undulations are marked with the dashed vertical lines.}
\label{fig:bumpsample}
\end{figure}

Our large sample of SLSNe-I and their LCs with excellent phase coverage provide a great opportunity to examine the LC undulation properties systematically. We perform analysis of the LCs of the 73 events in the gold and silver subclasses. All four events in the bronze class are excluded because of the sparse phase coverage of the LCs. To quantitatively identify the undulations, we compute the residual LC (RLC) by subtracting out a smooth baseline (produced by the {\tt MOSFiT} models) from the observed LC. 
In our modeling and analysis, we didn't manually exclude the undulating phases. The model LCs will go through undulations, and the resulting bumps and dips in the RLCs reflect how the LCs of SLSNe-I deviate from the standard models. %Note that identifying ``bumps''/``dips'' highly depends on the choice of the baselines, which can be ambiguous in many cases. But the magnitude changes between the bumps/dips are much less affected.
The observed data is interpolated using the GP regression. One example is shown in Figure~\ref{fig:bumpsample}, where the RLC is shown in the bottom panel including errors due to both the GP interpolation and the baseline model. In the RLC, the strength and phase of the bump/dip can be mathematically determined by their local maximum and minimum, marked as black dots in Figure~\ref{fig:bumpsample}. The maximum amplitude between the adjacent minimum and maximum, recorded as $\Delta$Mag$_{\rm RLC}^{\rm max}$, defines how much the LC undulates. The hatched area shows the time interval of the undulation, and the detailed properties of the LC undulations are discussed in \S\ref{subsec:Undulationproperty}.

One key element in computing the RLC is how to define the smooth baseline LC. Polynomial or GP regression fitting can produce smooth baselines, but they are also easily influenced by intrinsic bumps and dips. These two methods can in principle be applied to the rebinned LCs to smooth out the intrinsic LC variations. However, to achieve optimal results, both methods rely heavily on manual labor. They are not the best choice for our large sample. Instead, we adopt the best fit physical model LCs to define the smooth photometric evolution, derived by running {\tt MOSFIT} on the LCs (see \S\ref{sec:modeling} for details). SN\,2018bym can be equally well ($|\delta \chi^2| < 2\%$) fit by both the magnetar and CSM+Ni models, and we use the magnetar model as its baseline. %Here we summarize the LC fitting results from \S\ref{sec:modeling} again. Of the 70 SLSNe-I with sufficient data, 16 events are better fit by CSM+Ni model, 7 are better fit by magnetar model. The rest 47 events can be equally well fit by both models and we use the magnetar model as their smooth baselines.
The last three events (SN\,2018lzw, SN\,2018gkz and SN\,2019otl) are poorly sampled before peak phases, and their LCs do not show significant bump or dip features. Their baseline LCs are defined by a third-order polynomial fit. The significance of each RLC amplitude is set by its ${\rm SNR} = |\Delta$Mag$_{\rm RLC}^{\rm max}|/1\sigma$, where the $1\sigma$ error includes the uncertainty from {\tt MOSFiT} and GP interpolation. The significance of each undulation is determined in the same way.

The authenticity of the undulations is affected by both the photometric data and the model baselines. 
For the data part, to minimize the impact of occasional photometric outliers, we require the time separation between extrema to be $>5$\,days. And to avoid artificial bumps/dips produced by the interpolation in the absence of data, we require each extremum in the RLC to have at least two nearby data points within 10 days. 
For the model part: {\bf[1]} The choice of baselines ({\it i.e.} using the magnetar or CSM+Ni baseline) barely affects the identifications of undulating events, though detailed properties of undulations may change. All of the strongly undulating events and half of the weak ones identified below can pass the criteria and be identified as undulating events, no matter which model baseline is chosen. 
{\bf[2]} In occasional cases ({\it e.g.} SN\,2019lsq), the $g-$band model LC from the {\tt MOSFiT} is significantly lower than the observed LC at the peak, while the $r-$band model LC fits the observed one well. It is possible that the excess detected in $g$ band is real, however, it is also likely that the $g$-band luminosity at peak is underestimated by the models. This may be due to that the SED and the temperature evolution functions in {\tt MOSFiT} can not fully match the real ones. We exclude the undulations at the $g-$band peaks of SN\,2019lsq and SN\,2019neq in the following analysis, but still mark them in Figures~\ref{fig:phasedmag} and \ref{fig:bumpenergy} with solid points. There could be more undulations, {\it e.g.} in SN\,2019cdt and SN\,2019nhs, that are simply caused by the deficiency of the baseline models.  However, it is difficult to distinguish whether an undulation is intrinsic or due to the local failure of the model, especially when the model LC matches the observed one in most epochs. 
{\bf[3]} It is also possible that the smooth events ({\it e.g.} SN\,2020qef) based on the CSM+Ni model are actually powered by the magnetar model plus undulations, especially considering the fact that the kinetic energies of some CSM-favored events seem too large. This would lead to an underestimation of the fraction of SLSNe-I with undulations.

%To avoid artificial bumps/dips produced by the interpolation in the absence of data, we require each extremum in the RLC to have at least two nearby data points within 10 days. To minimize the impact of occasional photometric outliers, we require the time separation between extrema to be $>5$\,days. The significance of each RLC amplitude is set by its ${\rm SNR} = |\Delta$Mag$_{\rm RLC}^{\rm max}|/1\sigma$, where the $1\sigma$ error includes the uncertainty from {\tt MOSFiT}. The $\rm SNR$ value determines the significance of each undulation.

\begin{figure}[htp]
\includegraphics[width=0.48\textwidth]{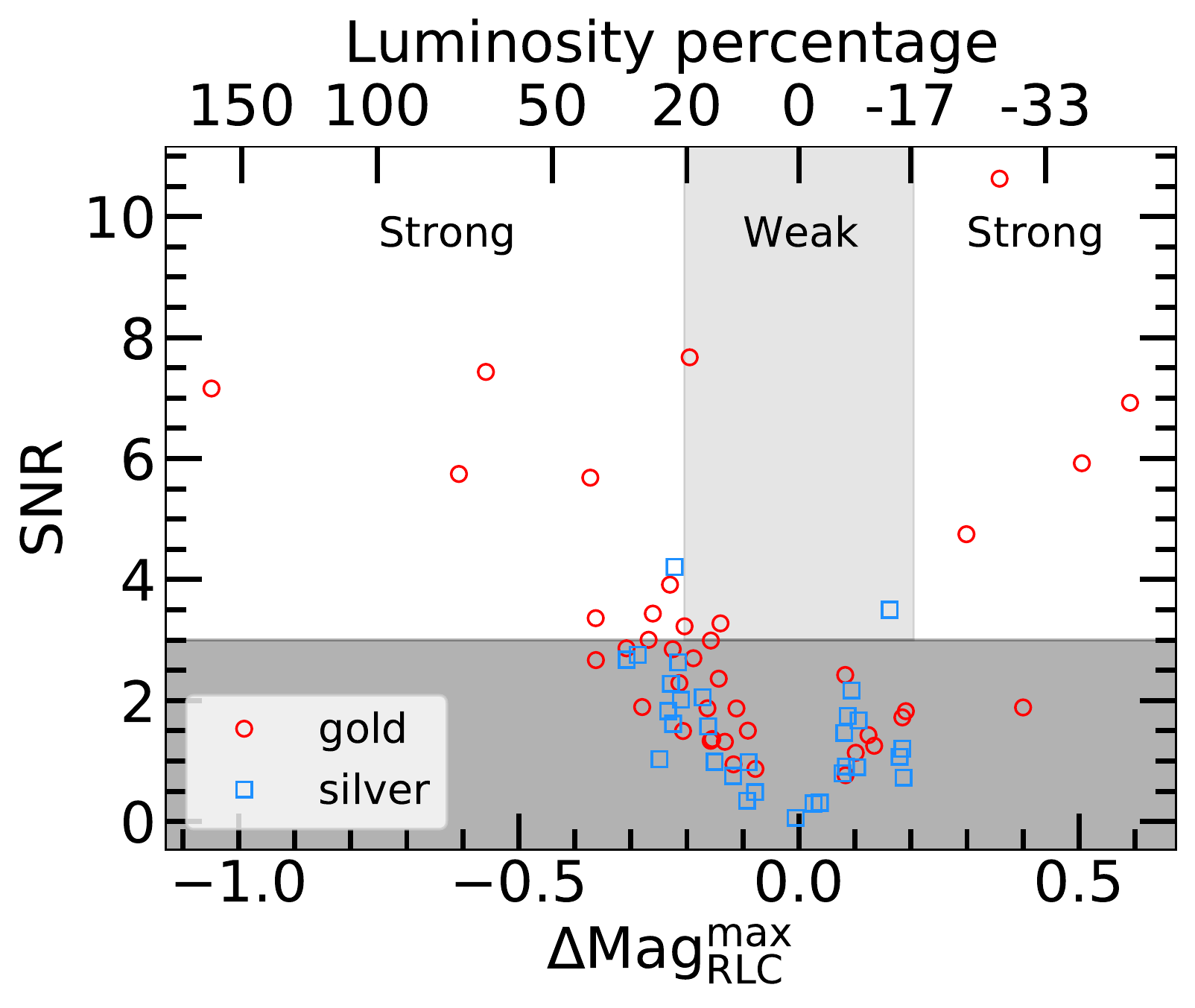}
\caption{The maximum amplitude of the undulation versus the significance. X-axis shows $\Delta$Mag$_{\rm RLC}^{\rm max}$, the maximum amplitude of the RLC (See Figure~\ref{fig:bumpsample}) and y-axis is the SNR, {\it i.e.} the amplitude divided by the uncertainty. 
The maximum amplitude is measured from either $g$- or $r$-band LCs depending on which one has the larger $|\Delta$Mag$_{\rm RLC}|$ and higher SNR.
The energy percentage of the bump/dip relative to the smooth baseline LC is computed and labeled at the top. The region of SNR $<3$ is marked as the dark grey region and the weak undulation amplitude region with $|\Delta$Mag$_{\rm RLC}^{\rm max}|<0.2$\,mag is in light grey region. Events in the gold and the silver class are plotted in circles and squares, respectively.}
\label{fig:bumpdistribution}
\end{figure}

\begin{figure*}[t!]
\includegraphics[width=\textwidth]{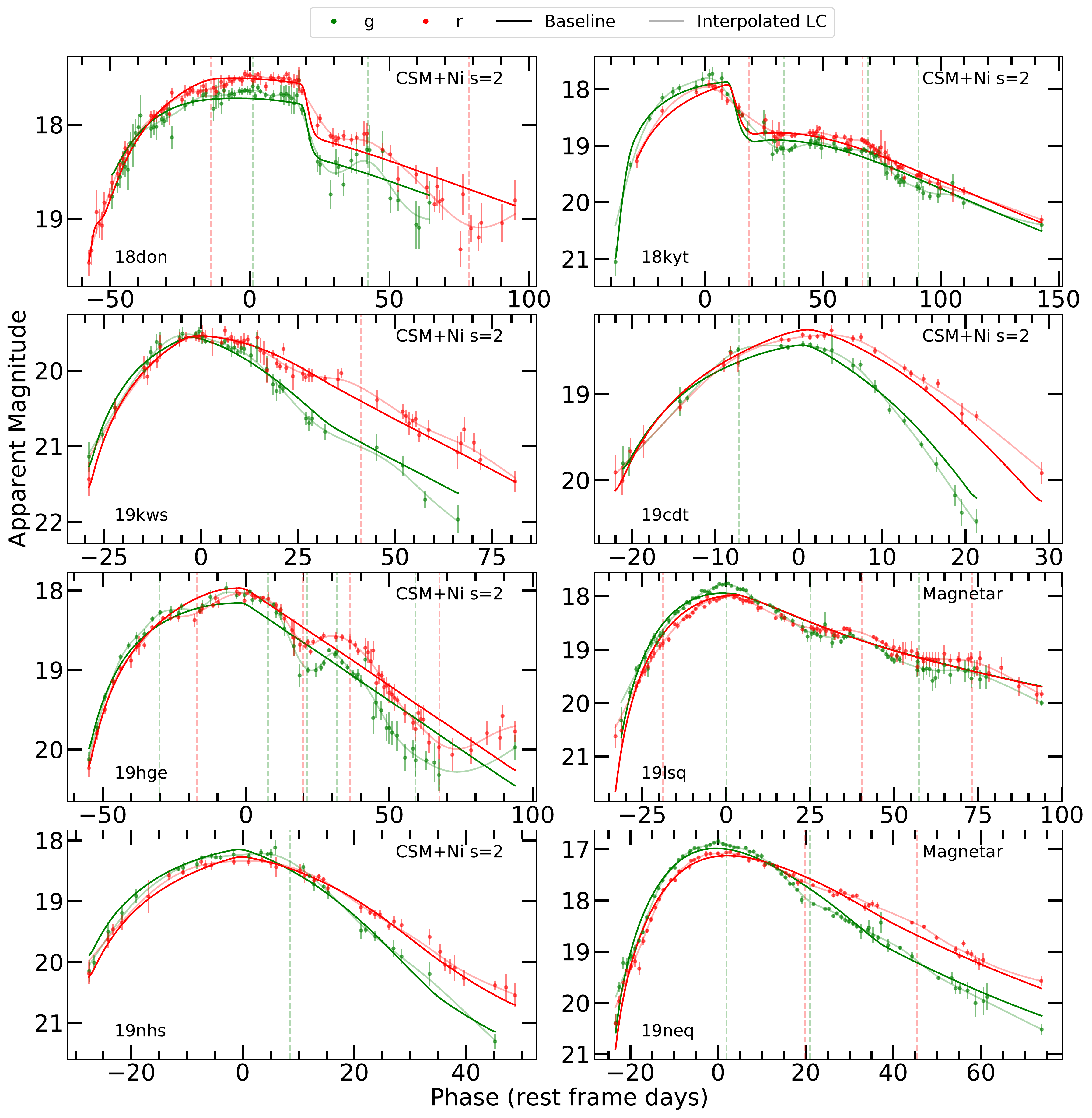}
\caption{SLSN-I sample with strong undulations. Baseline models and GP interpolated LCs are plotted in solid lines and translucent lines respectively. The models of the baselines are labeled at the top right corner. 
The peaks/dips of undulations are marked with the dashed vertical lines.
}
\label{fig:sigbump0}
\end{figure*}

\begin{figure*}[t!]
\includegraphics[width=\textwidth]{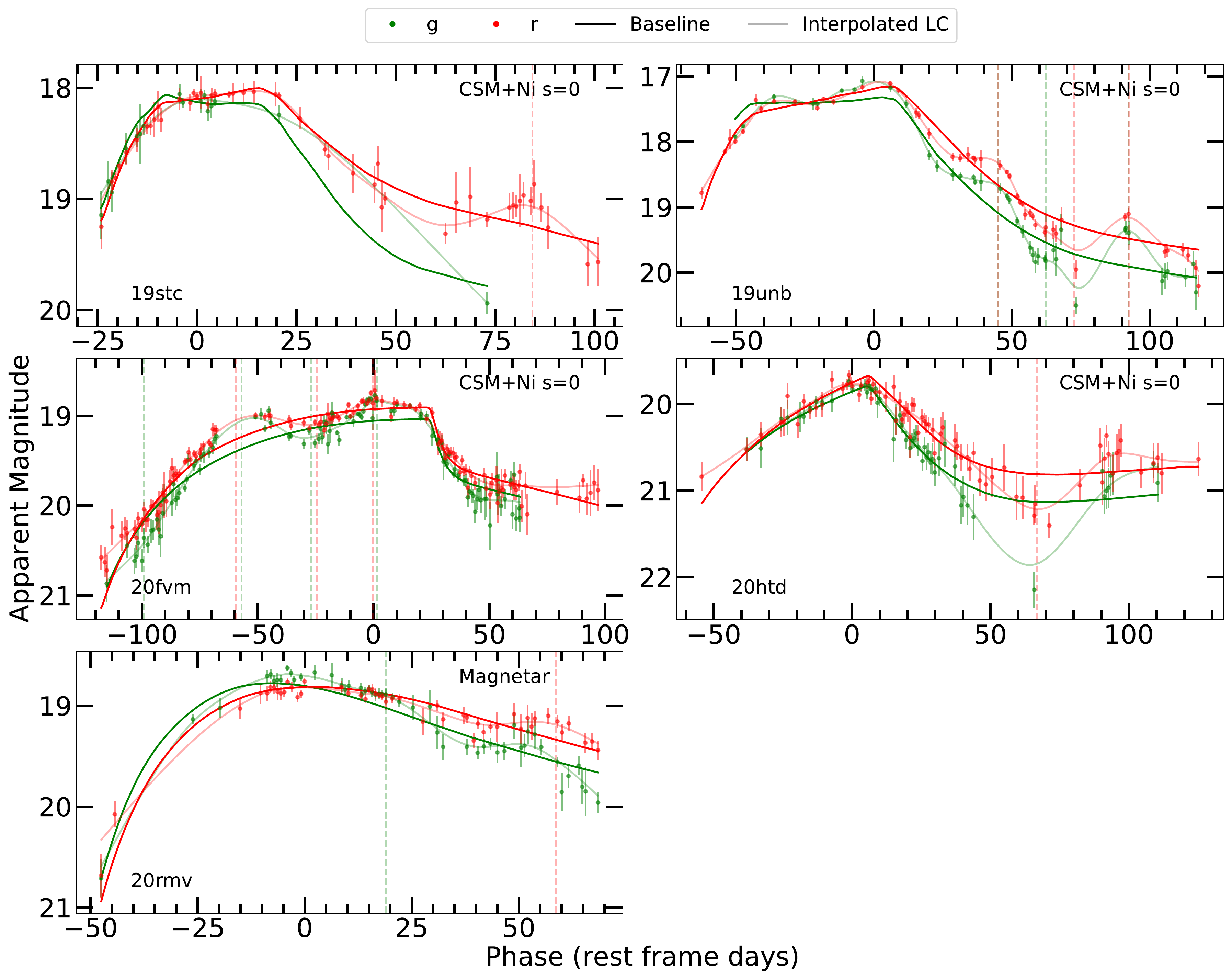}
\caption{Similar to Figure~\ref{fig:sigbump0}.}
\label{fig:sigbump1}
\end{figure*}

\begin{figure*}[htp]
\centering
\includegraphics[width=\textwidth]{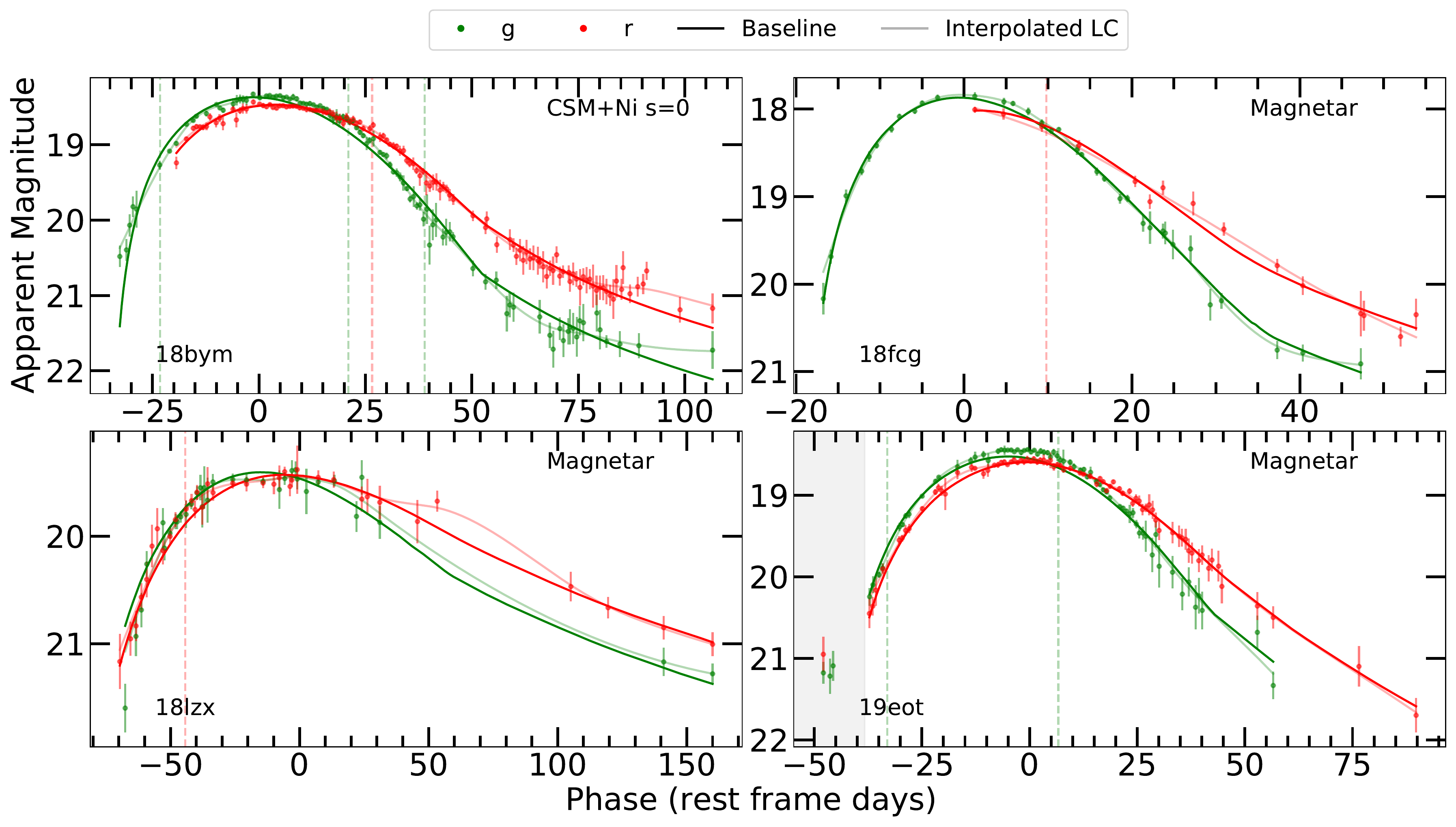}
\caption{SLSN-I sample with weak undulations. Similar to Figure~\ref{fig:sigbump0}. The grey shaded area marks the data excluded from the fitting.
}
\label{fig:weakbump}
\end{figure*}

Figure~\ref{fig:bumpdistribution} shows $\Delta$Mag$_{\rm RLC}^{\rm max}$ versus SNR. $\Delta$Mag$_{\rm RLC}^{\rm max}$ reflect the luminosity ratio between the bumps/dips relative to the baselines, so we also label the luminosity percentage at the top x-axis of Figure~\ref{fig:bumpdistribution}, where $+20\%$ and $-17\%$ correspond to $\pm 0.2$\,mag, respectively. We define a strong undulation as $\rm SNR \ge 3$ and $|\Delta$Mag$_{\rm RLC}^{\rm max}| \ge 0.2$\,mag, and a weak undulation as $\rm SNR \ge 3$ and $0.1\,{\rm mag} < |\Delta$Mag$_{\rm RLC}^{\rm max}| < 0.2$\,mag. A total of 17 events have LC undulations with $\rm SNR \ge 3$. Of these, 13 are strongly undulating sources and 4 weak ones. 
%SN\,2018hpq is manually moved from strongly undulating sources to weak ones due to lack of sufficient data around the undulation.
We list the strongly and weakly undulating events in Table~\ref{tab:undulatelist}. The LCs of strongly undulating sources are shown in Figures~\ref{fig:sigbump0} and \ref{fig:sigbump1}, and those of the 4 weakly undulating events are presented in Figure~\ref{fig:weakbump}.
It is worth noting that the CSM-favored fraction (62\%, 8/13) in strongly undulating events is significantly higher than that in our whole sample (23\%, 16/70). This is unlikely due to that the CSM+Ni model may create more artificial undulations, since all the strongly undulating events also show strong undulations based on their magnetar baselines. Thus, this indicates that the undulations are more likely to occur in the CSM-favored event, or that the undulations make the CSM+Ni model a better fit. 

Based on the above analysis, the fraction of undulating LCs is estimated to be $23\%$ (17/73). If counting only the strongly undulating events, this fraction is $18\%$ (13/73) while the fraction of weak undulations is $5\%$ (4/73). However, the LCs of the silver events usually do not have complete phase coverage and the undulation can be missed due to the lack of data. If we examine only the 40 events in the gold class, the fraction of undulating events is $25-52\%$ (15/40, CL$=95\%$), and the fraction of strong undulations is $18-44\%$ (12/40). This suggests that the LC undulations are very common in the SLSN-I population.
We note that sometimes it's difficult to distinguish weak undulations and observation biases, so we suggest using the fraction of the strongly undulating events in gold sample, {\it i.e.} $18-44\% $.
As discussed in the paragraph about the authenticity of undulations in this subsection, some undulations could be simply identified due to the deficiency of baseline models. This may apply to the undulations of SN\,2019cdt and SN\,2019nhs. Moreover, if removing these two events, the fraction of undulating events will be $14-39\%$ (10/40, CL$= 95\%$).

\Needspace{15\baselineskip}
\begin{center}
\begin{longtable}[htbp!]{ll|l}
%\begin{table}[htbp]
\caption{List of 17 undulating events}\\
%\centering
%\begin{tabular}{ll|l}
\toprule
\multicolumn{2}{c|}{Strong} & \multicolumn{1}{c}{Weak} \\
\midrule
SN\,2018don & SN\,2019neq & SN\,2018bym \\
SN\,2018kyt & SN\,2019stc & SN\,2018fcg \\
SN\,2019kws & SN\,2019unb & SN\,2018lzx $\star$ \\
SN\,2019cdt & SN\,2020fvm & SN\,2019eot \\
SN\,2019hge & SN\,2020htd &  \\
SN\,2019lsq & SN\,2020rmv $\star$ &  \\
SN\,2019nhs &  &  \\
\bottomrule
%\end{tabular}
%\begin{tabular}{p{5cm}p{1cm}<{\centering}p{1cm}<{\centering}}
\hspace*{\fill} \\
\multicolumn{3}{l}{$\star$ means the event in the silver sub-sample.}
%\end{tabular}
\label{tab:undulatelist}
%\end{table}
\end{longtable}
\end{center}

\subsection{Time scales and energetics of LC undulations}
\label{subsec:Undulationproperty}

From the RLCs, we can measure additional parameters, including phase, strength, time interval and energetics of each undulation. In this subsection, we focus on the 13 strongly undulating SLSNe-I, each of which may have multiple bumps/dips. We define the phase and the strength of each bump/dip by the time relative the peak phase and the amplitude of the RLC, Mag$_{\rm RLC}$, when the RLC reaches the maximum/minimum, respectively. Counting only the significant undulations with the RLC SNR $>3$\footnote{The undulations of SN\,2019kws, SN\,2019nhs, SN\,2020afag and SN\,2018don (one $g-$band undulation at $+42$\,days) are manually included, since their adjacent minimum and maximum each contribute to half of the $\Delta$Mag$_{\rm RLC}^{\rm max}$ and neither has SNR $>3$. }, we identify 23 undulations in the both $g$ and $r$ band. This implies that on average, each undulating LC has roughly 2 significant bumps and dips. 

\begin{figure}[htp]
\includegraphics[width=0.5\textwidth]{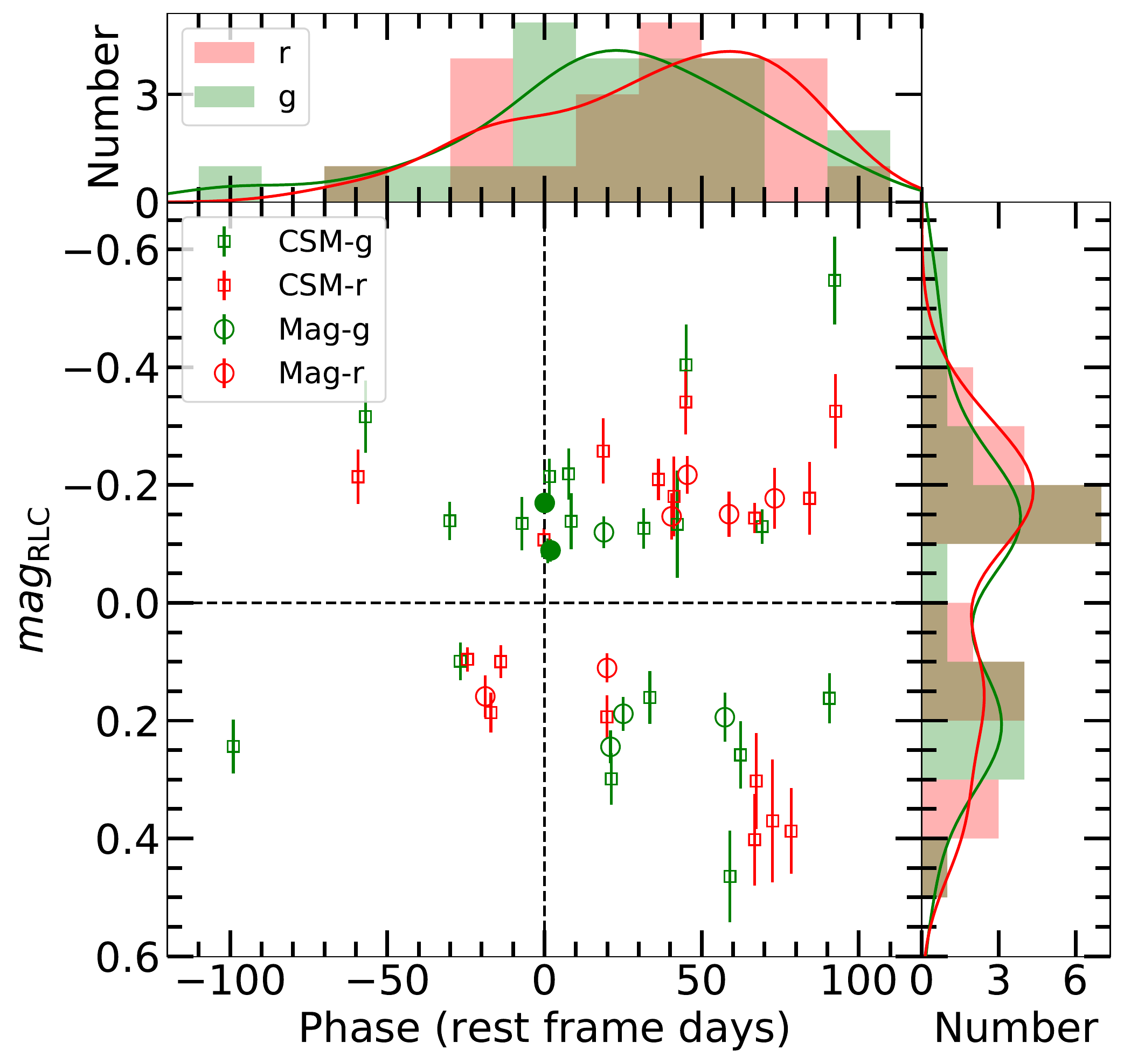}
\caption{The phases versus the strengths of undulations. The strength is defined by the magnitude of the residual LC, $\rm Mag_{RLC}$, where negative value means bumps. $g$- and $r$-band undulations are plotted in green and red, respectively. The events using the CSM+Ni and magnetar baselines are marked with squares and circles, respectively. The vertical black dashed line marks the peak phase and the horizontal one marks the boundary between bumps and dips. The histograms along x and y axes show the distributions of the phase and the strength, respectively. The solid lines show the kernel density estimation of the distributions. The undulations at the $g-$band peaks of SN\,2019lsq and SN\,2019neq, which could be due to the deficiency of the model, are marked with green solid points.}
\label{fig:phasedmag}
\end{figure}

Figure~\ref{fig:phasedmag} presents the phase against the strength of the significant undulations, as well as their distributions. Bumps and dips each account for around half of the undulations. $76\%$ of the undulations appear at post-peak phases and all the strong undulations ({\it e.g.} $|\rm Mag_{RLC}| > 0.3$\,mag) occur post peak. This indicates that pre-peak undulations are weaker and less common compared with post-peak ones, which is expected since the pre-peak LCs are usually much shorter than that of post-peak.

The next parameter is the time interval, $\Delta t$, which is defined as the time duration between two phases when their RLC values are zero (see Figure~\ref{fig:bumpsample}). The $\Delta t$ in strongly undulating events have a wide range from 11 to 61 days, with a median value and 1$\sigma$ error of $28.8^{+14.4}_{-9.1}$\,days. For the central engine models \citep[magnetar or black hole fallback,][]{Kasen2010, Dexter2013}, the LC undulations could originate from the temporal change of the central source (see discussion in \S\ref{sec:discuss}). However, any temporal variation from a central engine is smoothed by the photon diffusion in the expanding ejecta, and rapid variations would get washed out if the diffusion time is long.

Here $\Delta t$ can be compared with $t_{\rm diff}^{\rm eff}$, the effective photon diffusion time scale, defined as 
\begin{equation}
t_{\rm diff}^{\rm eff}=\left(  \frac{2 \kappa M_{\rm ej}}{\beta c V_{\rm ej}} \right)^{1/2},
\label{eq:tdiff}
\end{equation}
where $\beta$ is a constant that equals 13.7, and $\kappa, M_{\rm ej}$ and $V_{\rm ej}$ are opacity, ejecta mass, and velocity, respectively. The instantaneous diffusion time scale, $t_{\rm diff}$, can be calculated as $\left( t_{\rm diff}^{\rm eff} \right) ^2 /t$, where $t$ is the phase relative to the explosion date taken from the {\tt MOSFiT} modeling. After the peak, the ejecta gradually becomes transparent and the photospheric radius recedes inward. The instantaneous diffusion time of photons becomes shorter. The ratio of $\Delta t/t_{\rm diff}^{\rm eff}$ or $\Delta t/t_{\rm diff}$ (approximately the $\delta$ in \hyperlink{cite.Hosseinzadeh2022}{H22}) is a good indicator of whether the variable central engine scenario may drive the LC undulations. Specifically, if this ratio is less than 1, this model can not explain the undulations because of the smearing effect from the photon diffusion process.

\begin{figure}[htp]
\includegraphics[width=0.5\textwidth]{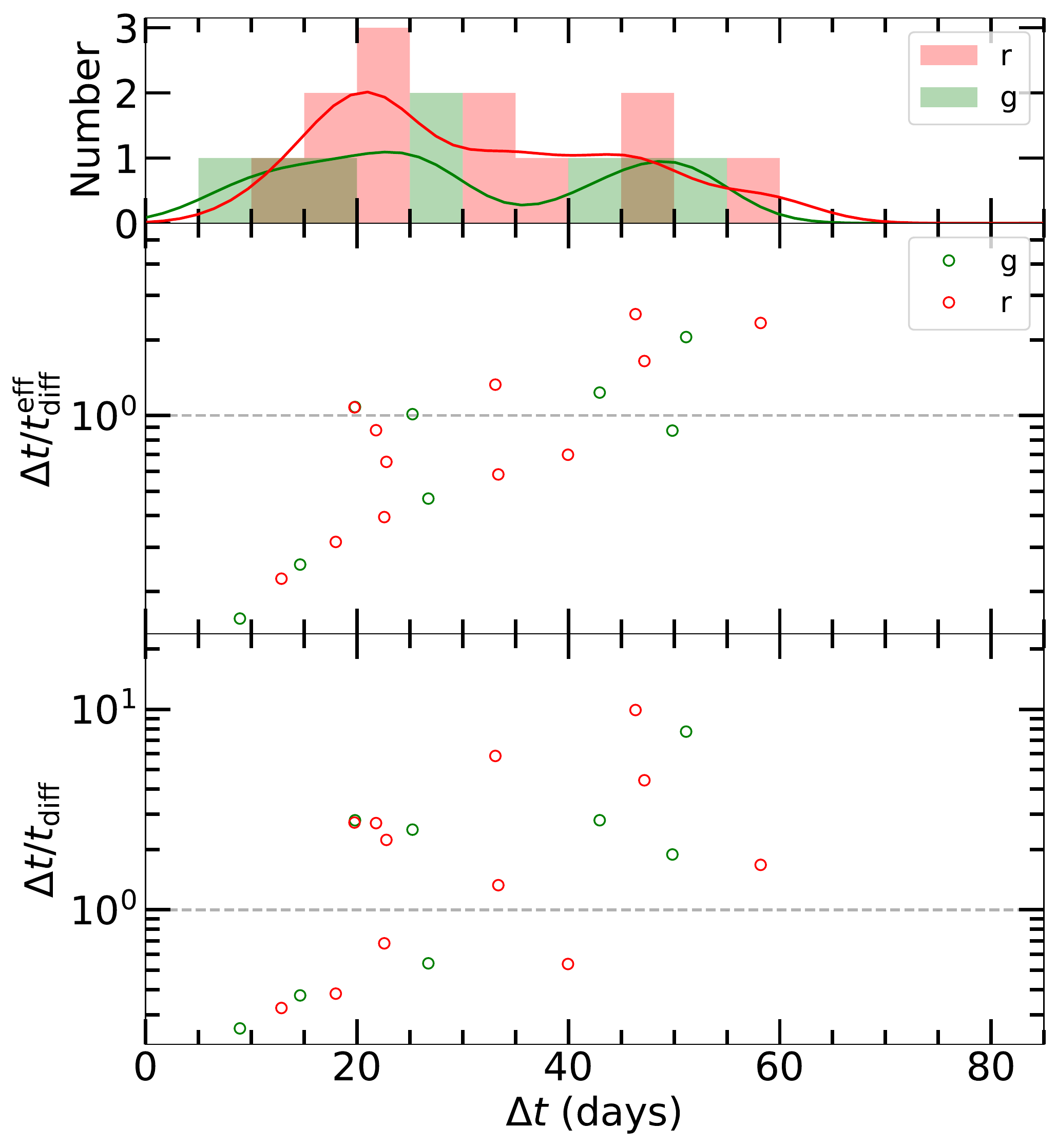}
\caption{Time durations of the undulations, $\Delta t$, measured from the RLCs of the 5 strongly undulating events with the magnetar baselines. The top panel shows the distribution of $\Delta t$ and the solid lines represent the kernel density estimations. The middle panel shows the ratio of $\Delta t$ and the effective diffusion time scale $t_{\rm diff}^{\rm eff}$. The bottom panel shows the ratio of $\Delta t$ and the instantaneous diffusion time scale $t_{\rm diff}$ at the corresponding undulation phase. The dashed horizontal lines mark $\Delta t = t_{\rm diff}^{\rm eff}$ and $t_{\rm diff}$.}
\label{fig:bumptime}
\end{figure}

Of the 13 strongly undulating events, only 5 can be fit by the magnetar model. Figure~\ref{fig:bumptime} displays their $\Delta t$ versus the ratios of $\Delta t/t_{\rm diff}^{\rm eff}$ and $\Delta t/t_{\rm diff}$. We find that $55\%$ (11/20) of the undulations are shorter than $t_{\rm diff}^{\rm eff}$ and $35\%$ (7/20) are shorter than $t_{\rm diff}$. $60\%$ (3/5) and $20\%$ (1/5) of the strongly undulating events have shorter undulations relative to either $t_{\rm diff}^{\rm eff}$ or $t_{\rm diff}$, respectively. Roughly half of the LC undulations ($20-60\%$) have time intervals shorter than the photon diffusion time scales. This implies that emission variations of the central engine could be a viable physical explanation for about $50\%$ of the undulations. For the other $50\%$, other physical processes are needed because the short time scale undulations would get smoothed out by photon diffusion. 

\begin{figure}[htp]
\includegraphics[width=0.5\textwidth]{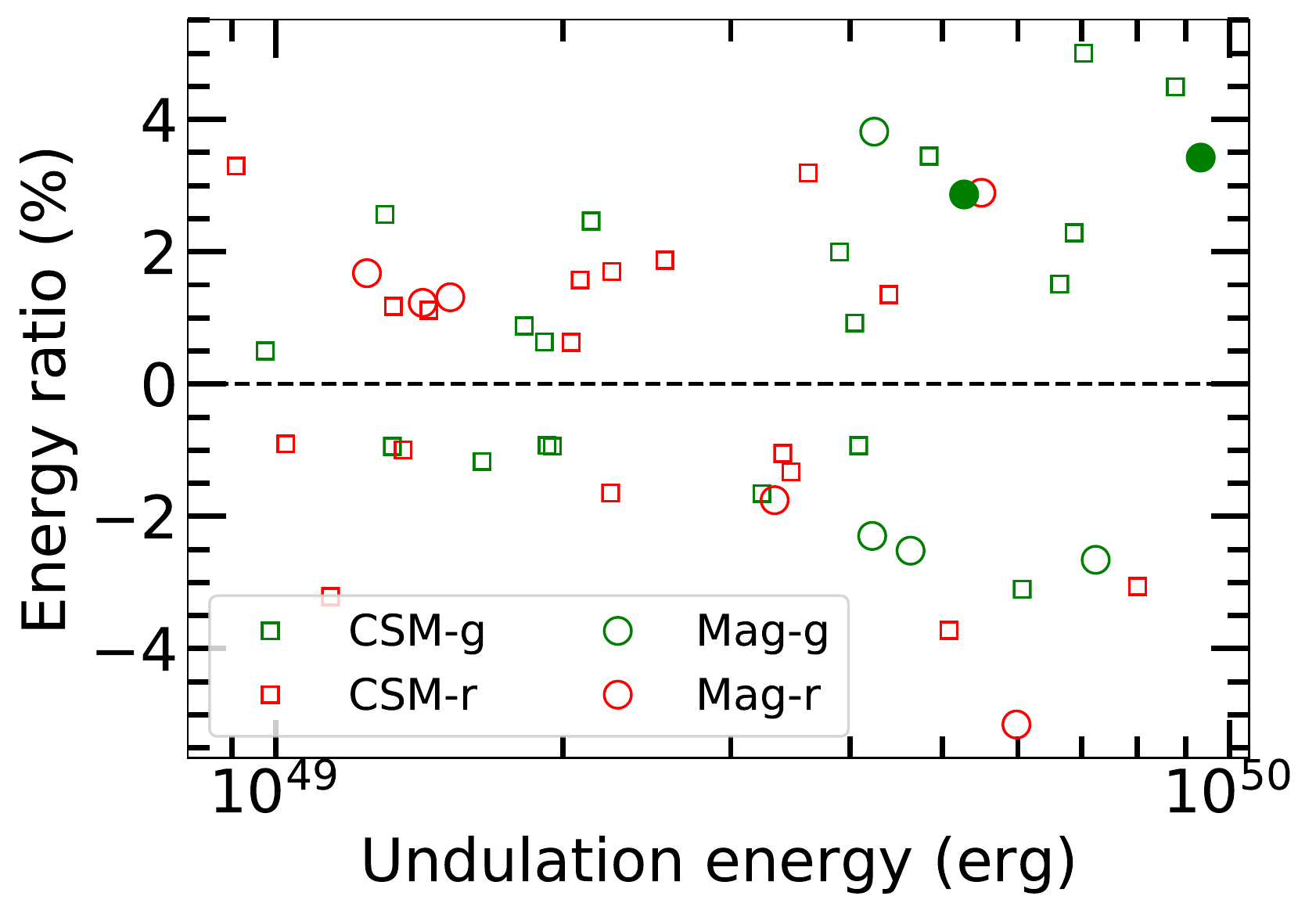}
\caption{Energetics of the bumps/dips in the 13 strongly undulating events. X-axis is the integrated energy over the time interval of the undulation. Y-axis shows the ratio of undulation energy to the total radiative energy in $g$ or $r$ band over the entire LC. The undulations at the $g-$band peaks of SN\,2019lsq and SN\,2019neq, which could be due to the deficiency of the model, are marked with green solid points.}
\label{fig:bumpenergy}
\end{figure}

Finally, we measure the monochromatic energy of each significant undulation, $\rm E_{\lambda,\rm undu}$, by integrating the flux differences between the LC and its baseline model (shown as the hatched area in Figure~\ref{fig:bumpsample}). We compute the ratio between $\rm E_{\lambda,undu}$ and the total monochromatic energy of the entire LC, $\rm E_{\lambda,total}$ for each undulation. Figure~\ref{fig:bumpenergy} plots the undulation energy versus the ratio. It is worth noting that most undulations appear to be quite energetic, with absolute values between $9.1\times10^{48}$ to $8.8\times10^{49}$\,erg. However, they constitute only a small fraction of the total radiative energy, with the median energy ratio (absolute value) of $1.7\%^{+1.5\%}_{-0.7\%}$.  %The outliers in  have very high energy ratios, and these SLSNe-I, {\it e.g.} SN\,2019stc and SN\,2020auv, have bright secondary peaks. The physical drivers powering these outliers may be different from the mechanisms working for the majority of the weaker undulations (see discussion below).

\subsection{Correlations}
\label{subsec:correlation}

The undulation properties may have intrinsic correlations which can reveal the possible physical mechanism. We test for correlations between the phases (relative to explosion), absolute strengths, energies, time intervals of undulations and %the half-maximum rise time timescales $t_{rise}$ (the $t_{rise,1/2}$ from Paper I). 
the rise time $t_{\rm rise}$ (the time interval between explosion and the LC peak). 
Here the explosion date is determined by {\tt MOSFiT}.

\hyperlink{cite.Hosseinzadeh2022}{H22} claimed that the phases of the post-peak bumps are moderately correlated with the LC rise times, implying that such bumps tend to happen at a certain evolutionary stage. However, observational selection effects play a significant role in their result because SLSNe-I with long rising time scales are also slow declining, and undulations are preferentially detected in slow fading events. It would be difficult to observe late time undulations in rapidly evolving events.

\begin{figure}[htp]
\includegraphics[width=0.5\textwidth]{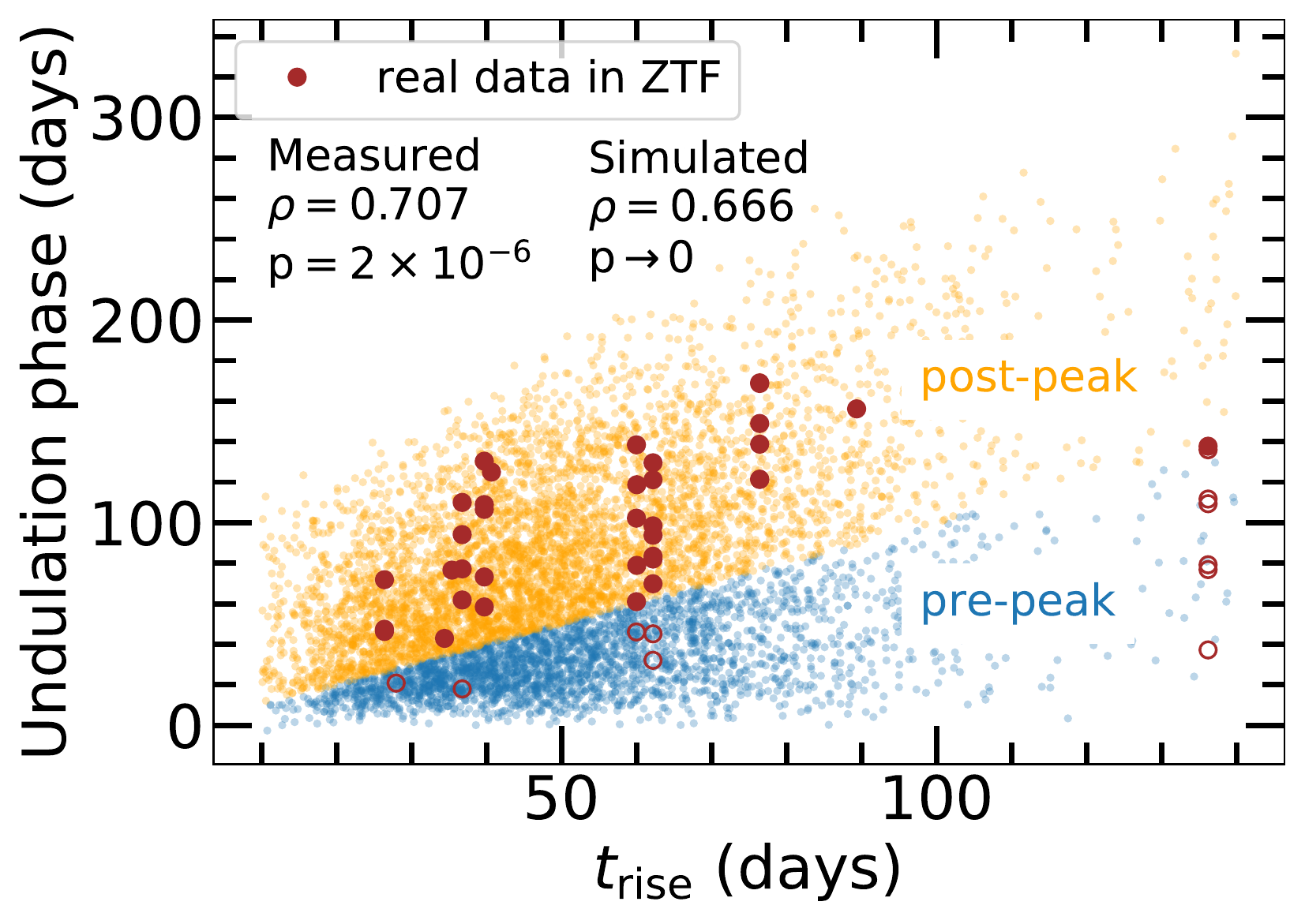}
\caption{The observational selection effects on the rise times and the undulation phases (relative to explosion). Pre- and post-peak simulated undulations (1\% of total) are plotted here with blue and yellow dots, respectively. The measured data from our sample is presented in brown, with open/solid dots for pre/post-peak bumps. 
By assuming undulations randomly occur in LCs, we find that observational effects will bring a strong correlation ($\rho \approx 0.67$), which is comparable to the value ($\rho \approx 0.5$) measured in \hyperlink{cite.Hosseinzadeh2022}{H22} and the one measured in our sample ($\rho=0.71,~ p = 2 \times 10^{-6}$), illustrating that such a correlation is mainly caused by an observational bias.}
\label{fig:ObEtest}
\end{figure}

We carry out a simulation to quantify this observational bias. %First we set $N=10^4$ SLSN-I events with random $t_{\rm rise}$ from 10 to 85 days (approximate the range of our sample).
First we set $N=10^6$ SLSN-I events, the $t_{\rm rise}$ of which follows the same distribution function as our whole SLSN-I sample ({\it i.e.} centered at $\sim40$\,days with an extended tail to $\sim140$\,days). 
Second we let the undulations {\bf randomly} occur in a time range from explosion to a maximum detectable time $t_{\rm max}$. The $t_{\rm max}$ usually highly correlates with the $t_{\rm rise}$ ($\rho=0.60,~ p < 10^{-7}$ in our sample), since $t_{\rm max} \approx t_{\rm rise} + t_{\rm decay}$ and it has been shown that slow-rising SLSNe-I tend to decay slow by different people \citep[Paper I;][]{Nicholl2015b,DeCia2018}. The $t_{\rm max}$ is thus set random in a $\pm 67$-day-wide range along the empirical relation derived from our sample, $t_{\rm max} \approx 1.9~ t_{\rm rise} + 35$\,days. As shown in Figure~\ref{fig:ObEtest}, if only the post-peak undulations are taken into account like \hyperlink{cite.Hosseinzadeh2022}{H22}, the observational bias introduces a strong correlation ($\rho \approx 0.67,~ p \to 0$\, since the number of simulated events is large) between the undulation phases and the $t_{\rm rise}$. This is comparable to the value ($\rho \approx 0.5, p \approx 0.01$) measured by \hyperlink{cite.Hosseinzadeh2022}{H22} and the one ($\rho=0.71,~ p = 2 \times 10^{-6}$) measured with our real data. 
We further simulate samples with Gaussian or flat distributed $t_{\rm rise}$, and adjust the parameters and the random range of the empirical relation in a wide range from 50\%\ to two times. This correlation always exists ($\rho = 0.53 - 0.78$). 
Even if we calculate the correlation coefficient using both pre- and post-peak undulations, the measured value ($\rho=0.48,~ p<10^{-4}$) and the one caused by observational bias ($\rho \approx 0.37$) are still comparable. Note that the undulations with the longest $t_{\rm rise}$ are from SN\,2020fvm, which has two LC peaks and an unusually long rise time ($136$\,days). No matter whether we use the shorter $t_{\rm rise}$ calculated using its first but fainter peak or simply delete these points, the correlation does not change. 
We conclude that the correlation between the undulation phase and the rise time is very likely due to observational selection effects and not a physical relation. 

\begin{figure}[t]
\includegraphics[width=0.5\textwidth]{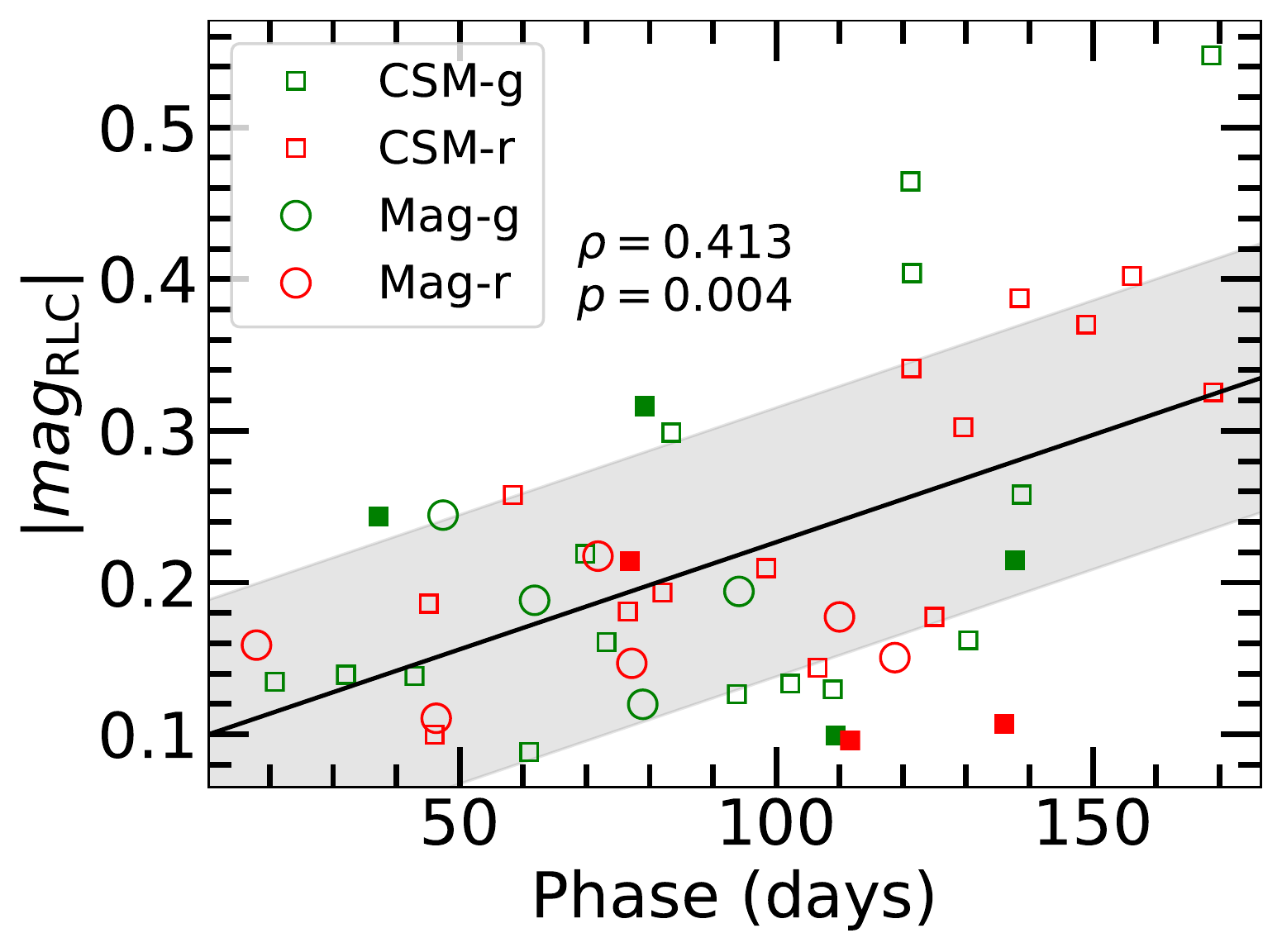}
\caption{The correlation between the absolute strengths and phases of undulations. The correlation coefficient and the $p$-value are listed in the figure. The correlation will be much stronger if we remove the data from one unusually slow-evolving event SN\,2020fvm. The data from SN\,2020fvm are marked with solid symbols. The black solid line and the grey region show the result and 1$\sigma$ error of the linear fit.}
\label{fig:correlation}
\end{figure}

%For other parameters, we do not find significant correlations which can reveal any possible physical mechanism. 
For other parameters, we only find weak positive correlation ($\rho=0.41,~ p=0.004$) between the phase and absolute strength as shown in Figure~\ref{fig:correlation}. If we exclude the specially slow-evolving event, SN\,2020fvm, 
%with an extremely strong and long-lived secondary peak, 
the correlation becomes much stronger ($\rho=0.58,~ p = 10^{-4}$). %The weak correlation may be biased because undulations at later phases are harder to detect among late time LCs with low brightness and poorer sampling. 
However, this correlation could also be affected by observational bias, since {\bf[1]} the errors of late time data are usually large which makes weak undulations hard to detect; {\bf[2]} the early time LCs are brighter and the undulations require more energy to reach the same magnitude strength. No significant correlation between the phase and undulation energy also proves this.  Our data are not good enough to investigate this further.

\subsection{Optical colors at bump phases}
\label{subsec:Undulationcolor}

\begin{figure*}[htp]
\includegraphics[width=\textwidth]{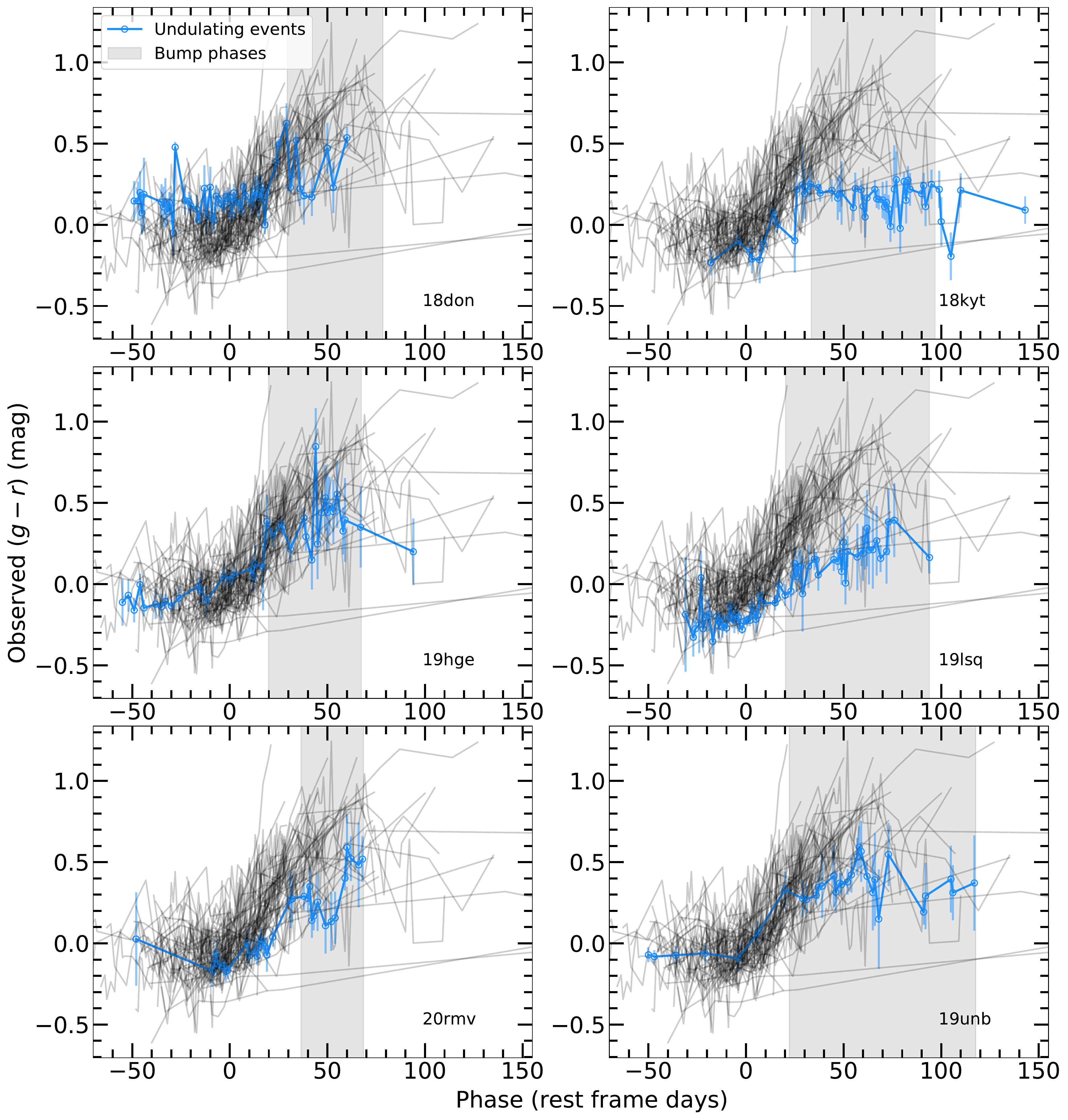}
\caption{The observed $(g-r)$ colors as a function of rest-frame days relative to the peak phases. The blue tracks in the six panels are for the 6 strongly undulating events, overlaid on top of the color tracks of the full sample (gray lines). The bump phases are marked by the shaded vertical bars. All 6 panels show that the observed $g-r$ colors turn bluer during the bump phases. }
\label{fig:bumpcolor}
\end{figure*}

The transient colors during the undulation phases could be an indicator of the physical processes. The observed $(g-r)$ colors of most of the undulations follow the general trend of the sample. However, six strongly undulating events stand out. In Figure~\ref{fig:bumpcolor}, the blue lines are the observed $(g-r)$ color evolution tracks of six strongly undulating events. For comparison, the observed $(g-r)$ colors of the rest of the sample are shown in grey (see details in Paper I). Here we mark with the shaded vertical bars the period of times when the LC undulations are in excess, {\it i.e.} bump phases. These are quantitatively defined by the time intervals between the minima in the RLCs. From Figure~\ref{fig:bumpcolor}, we note that during the bump phases, the observed $(g-r)$ colors are significantly bluer than that of the comparison sources. In particular, three events, SN\,2018don, SN\,2019hge and SN\,2020rmv, show much bluer colors when their LCs are varying, and turn redder again after or at the end of the bumps. The other three events, SN\,2018kyt, SN\,2019lsq and SN\,2019unb, are found to show much bluer and more stable colors than the general trend seen at late times ($>+40$\,days). This result suggests that the CSM interaction could be an important energy source, and can naturally explain both the blue colors and the strong secondary peaks at late times for these six events. The duration of the blue color phase may be affected by the thickness of the CSM. However, magnetar heating can not be completely ruled out as different magnetar deposition profiles could also impact the color evolution at late times \citep{Dessart2019}.

\subsection{LC undulations in Helium-rich SLSNe-Ib}
\label{subsec:undulateIb}

\citet{Yan2020} reported six He-rich SLSN-Ib events from the ZTF Phase-I SLSN-I sample. One additional event, SN\,2020qef, was thereafter spectroscopically classified as a SLSN-Ib by \citet{Terreran2020}. Of these seven He-rich SLSNe-Ib, five events, namely SN\,2018kyt, SN\,2019kws, SN\,2019hge, SN\,2019unb and SN\,2020qef, have gold LCs. Of these five SLSNe-Ib, four\footnote{SN\,2018kyt, SN\,2019kws, SN\,2019hge and SN\,2019unb} have strong undulations, three\footnote{SN\,2018kyt, SN\,2019hge and SN\,2019unb} show much bluer color during their bump phases and three\footnote{SN\,2018kyt, SN\,2019unb and SN\,2020qef} strongly prefer the CSM+Ni model over the magnetar model, as shown in Figures~\ref{fig:sigbump0}, \ref{fig:bumpcolor} and \ref{fig:CSM}. All the five well-sampled He-rich SLSNe-Ib have either strongly undulating LCs or the LCs are much better fit by the CSM+Ni model. This small sample appears to have a much higher undulation fraction and a higher fraction of CMS+Ni powered LCs than those of the full sample.  

These results suggest that CSM are present in He-rich SLSNe-Ib and leave significant imprints on their LCs. This is consistent with a scenario proposed in \citet{Yan2020}, where the progenitors of SLSNe-Ib have lost most of their hydrogen envelopes but have not had enough time to also lose {\bf all} of their helium layers. Because of the short time interval between the mass loss and the supernova explosion, it is likely that CSM are present near the progenitor stars. This scenario can naturally explain many of the observed characteristics of SLSNe-Ib, including He-rich spectra, LC undulations and blue colors during the bumps.

As for the remaining two of the seven SLSNe-Ib, SN\,2019obk and SN\,2019gam, both have poorly sampled LCs. The absence of undulations in their LCs could simply be due to lack of data. 

\section{Discussion} \label{sec:discuss}

We modelled and analyzed the LCs of 70 gold and silver SLSN-I events presented in Paper I. Based on two commonly used SLSN models, the magnetar model and the CSM+Ni model, we explore the properties and possible mechanisms which drive LC undulations.

\subsection{What drives LC undulations among SLSNe-I?}
\label{subsec:whyundulate}

One major finding is that LC undulations are common, about $18-44\% (12/40)$ of the SLSNe-I show such features. This fraction should be a lower limit as we count only undulations with strength $>0.2$\,mag and $\rm SNR > 3$ in gold sample, and some events are not well sampled at late times. 
%This result is consistent with the fraction of $44 - 76\%$ found in \hyperlink{cite.Hosseinzadeh2022}{H22}, which analysed the LC undulations based on a sample of 34 published SLSNe-I from multiple papers. Although the \hyperlink{cite.Hosseinzadeh2022}{H22} sample size is small, their LCs have good phase coverage out to $+100$\,days post-peak. 
The undulation fraction for SLSNe-I is therefore quite high, and likely also higher than that of SNe\,Ic \citep{Prentice2016,Lyman2016}, although the actual undulation fraction for SNe\,Ic has not been measured and requires future work. For comparison, the undulation fraction for SNe\,IIn appears to be quite low, only $1.4^{+14.6}_{-1.0}\%$ from a study of 42 events from Palomar Transient Factory \citep{Nyholm2020}. This result has large uncertainties, and is therefore worth further validation with better LCs from ZTF.

The important question is what physical processes are driving the observed LC variations. There are two main possibilities. First, in the central engine scenario, the power output of the central source -- either a magnetar or black hole fall-back accretion -- may have a temporal variation. This intrinsic variation will be modulated (or smoothed) by photon diffusion in the ejecta. At a given phase, variations shorter than the photon diffusion time scales will be smoothed out and not observable at that phase. Figure~\ref{fig:bumptime} in \S\ref{subsec:Undulationproperty} compares the undulation and photon diffusion time scales, illustrating that the central engine temporal variations could be a viable physical explanation for only $50\%$ of the undulations. For the other $50\%$, which have shorter time scale undulations, different mechanisms are likely at work.  

In the variable central engine scenario, as SNe evolve, the ejecta gradually becomes transparent. This implies that the LC undulations should be stronger and more easily to be observed at later times. Indeed, we find 76\% of undulations and several strongest ones all occur post peak (\S\ref{subsec:Undulationproperty}). The variable central engine scenario could be supported by the observed weak correlation between the phase and absolute strength of undulations, discussed in \S\ref{subsec:correlation}

It is also possible that the central energy output is constant, but the ejecta opacity may undergo temporal changes, which in turn causes the variations in photospheric emission. This idea was proposed to explain the LC undulation in the luminous transient ASASSN-15lh \citep{Margutti2017}. If a central energy source can increase the ionization of the ejecta, this can lead to higher optical opacity due to electron scattering. Furthermore, the UV opacity, dominated by metal line transitions, can decrease as the metal ions have fewer bound-bound transitions, leading to less opacity and higher UV emission. This may be the explanation for the extraordinarily UV bright, slowly evolving SLSN-I SN\,2019szu (see Paper I). In addition, when ejecta temperatures cool down, recombination of ions can lead to reduction of optical opacity. If this opacity decrease occurs quickly, it can manifest itself as a LC undulation. Another important point about the magnetar driven model is that our current understanding is still limited and more detailed 3D hydrodynamic simulations are beginning to find interesting results, such as strong instabilities and mixing in a magnetar-powered SN with CSM \citep{Chen-woosley2020}.  

The second mechanism is ejecta-CSM interaction. This process could be at work for at least $\sim$50\%\ of the SLSN-I undulations, and is further supported by the conclusion in \S~\ref{subsec:Undulation} that the undulations are more likely to occur in the CSM-favored events. CSI is an effective means to convert mechanical energy into thermal emission, and the CSM could have a variety of geometric and density distributions, {\it e.g.} shells or clumps. \citet{Vreeswijk2017,Liu2017b,Li2020} have applied the CSI model to explain the undulations observed in several SLSNe-I. The bluer $(g-r)$ colors during the bump phases observed among some of our SLSNe-I provide evidence supporting the CSM model (see Figure~\ref{fig:bumpcolor}). The simplistic picture is that the CSI can heat the ejecta and lead to both bluer colors and excess emission. 

The LC shape under the CSI mechanism is highly dependent on the density profile of the CSM. The events with multiple peaks could have multiple CSM shells or clumpy CSM structures. One possible explanation is that the progenitor undergoes violent episodic mass losses ({\it e.g.} PPISN). The other is that the SLSN-I progenitor not only has an extended CSM due to significant mass loss prior to the explosion, but also has a binary companion which sweeps up and enhances the CSM density while orbiting the progenitor. This idea was proposed for the radio LC of SN\,2001ig by \citet{Ryder2004}. 

Shock breakout has also been proposed to explain SLSN-I LC bumps at very early phases (so called double-peaked LCs). The energy source could be either magnetar/blackhole \citep{Kasen2016} or CSM interaction \citep{Moriya2012,Piro2015}.  Some of the basic ideas may be viable for explaining the LC undulations at late times, such as changing of ionization states thus opacities. However, it is not clear how these models can work for undulations at post-peak phases. More quantitative modeling is needed. 

Besides these major models, there are other possible scenarios. For example, \citet{Kaplan2020} suggests that the undulating LC of SN\,2018don \citep[][]{Lunnan2020} is the geometric effect of observing a different amount of emitting area from the two expanding photospheres of fast (polar) and slow (equatorial) outflows (or jets). The \citet{Kaplan2020} study is based on the results from \citet{Quataert2019}, which finds that the outer convection zones in yellow and red supergiants can generate enough angular momentum to form an accretion disk around the black hole. This model also predicts that these accretion flows could be highly time variable.

In our analysis, about half of the undulations are dips. Dips can actually occur in the LCs when the magnetar power or the CSM density goes down, or the opacity increases \citep{Moriya2012}. But identifying bumps/dips depends highly on the choice of baselines, which can be ambiguous in many cases. Bumps/dips in this work reflect more on how the observed LCs deviate from the standard models, and can not be clearly identified unless more accurate modeling is applied.
%Some of the bumps/dips can be the result of an under/overestimated baseline, but they still reflect the degree of undulations.

\subsection{Prevalence of H-poor CSM in SLSNe-I}
\label{subsec:rolecsm}
Of the 70 LCs, 16 have distinct features which are much better modelled by CSI with an assumed density profile (wind or constant). The fraction of SLSNe-I with H-poor CSM is between $25-44\%$, if we include events with undulations and smooth LCs preferentially fit by the CSI models.

This result has several implications for our understanding of the nature of SLSNe-I. First, if CSI plays an important role in driving undulations, such a high fraction implies that at the time of supernova explosion, H-poor CSM is likely present for a large fraction of SLSNe-I. This is the first time we have quantified how important CSI is for SLSNe-I using a carefully selected, large sample. Previously, many studies have preferred magnetar models for the SLSN-I population \citep{Inserra_2013, Nicholl2017b}. One reason is its simplicity. Another indirect reason is lack of observational signatures of CSM interaction. 

The presence of H-poor CSM around SLSNe-I also implies that the massive progenitor stars have not had enough time to completely disperse all of the outer envelopes before the core collapse happened. A fraction of these stars will have lost almost all their H-envelope but their He-rich outer layers are still present before the SN explosion. These events will appear as He-rich SLSNe-Ib \citep{Yan2020}. This also naturally explains why most of the SLSNe-Ib have undulating LCs which are better fit by the CSM+Ni model.

Furthermore, if the LC undulations could be explained by ejecta running into discrete CSM shells, this would imply that the progenitor mass loss is violent enough to eject large amounts of material. The CSM shell radius when the interaction occurs can be estimated by 
\begin{equation}
R \approx V_{\rm ej}\,t
\end{equation}
where $V_{\rm ej}$ is the ejecta velocity output from {\tt MOSFiT} and $t$ is the phase of the undulation. For the strongly undulating events in our sample, the radii vary from $1.3\times10^{15}$ to $1.3\times10^{16}$\,cm with a median value of $5.4\times10^{15}$\,cm. Assuming a stellar wind velocity of $10^2 - 10^3\,{\rm km\,s^{-1}}$, the CSM should be ejected several months to several decades before explosion.

\subsection{Comparison with {\rm H22}}
\label{subsec:H22}

Independent of our work, \hyperlink{cite.Hosseinzadeh2022}{H22} recently analysed the bumps in the post-peak LCs using a sample of 34 published SLSNe-I, which have LC phase coverage out to $+100$\,days post-peak. Eight events\footnote{SN\,2018bym, SN\,2018fcg, SN\,2018kyt, SN\,2019hge, SN\,2019lsq, SN\,2019neq, SN\,2019ujb and SN\,2019unb} are included in both their work and ours. Compared with \hyperlink{cite.Hosseinzadeh2022}{H22}, the greatest advantage of our sample is that our data is mainly from one single survey (ZTF), which is less affected by the systematic offsets and target selection from different telescopes. On the other hand, around half of our sample are not sampled up to $+100$\,days post-peak, and some undulations at such late phases will be missed in our results.

Both works use the magnetar model implemented in {\tt MOSFiT} and analyse LC undulations in the residual LCs by subtracting modeling LCs from the observed ones. We additionally fit the CSM+Ni model and choose the better one as our baselines. \hyperlink{cite.Hosseinzadeh2022}{H22} visually identified the bumps in LCs and masked the data during the bumps when modeling the baselines. To avoid human intervention in the selection of the undulating area, we keep all observed data in most events. We identify undulations by certain criteria of $\rm SNR>3$ and $|\Delta$Mag$_{\rm RLC}^{\rm max}|>0.2$\,mag.

Our conclusion for undulation fraction in SLSNe-I is $18-44\% (12/40)$, which is based on strongly undulating events in gold sample. This result seems slightly lower than the definite bump fraction of $44\% (15/34)$ measured by \hyperlink{cite.Hosseinzadeh2022}{H22}, but consistent within $1\sigma$. The difference could be due to various reasons, including: {\bf[1]} The baseline models and the model priors are different. Using the magnetar baseline may erroneously create undulations in the CSM-favored events while we could also missed the undulations when using the CSM+Ni model in the magnetar driven events; {\bf[2]} Undulations are identified via certain criteria in this work but via visual inspection in \hyperlink{cite.Hosseinzadeh2022}{H22}; {\bf[3]} Some events in this work are not well sampled at late time and thus late-time undulations can be missed. Among the eight events that are common in both works, the identification of four events\footnote{SN\,2018kyt, SN\,2019hge, SN\,2019neq, SN\,2019unb} are the same, while the rest four\footnote{SN\,2018bym, SN\,2018fcg, SN\,2019ujb, SN\,2019lsq} are different. More accurate modeling is needed for them. But in general, both work illustrate that LC undulations are common in SLSNe-I.

\hyperlink{cite.Hosseinzadeh2022}{H22} claimed that the phases of the bumps are moderately correlated with the LC rise times. We prove that such correlation is likely to be nonphysical, and is actually the result of observational selection effects. The energy source of LC undulations is still uncertain in both works, which requires more accurate models for SLSNe-I.

\section{Summary}
\label{sec:summary}
The three major results from our analysis are as follows.

\begin{enumerate}

\item LC undulations appear to be common, with $18-44\% $ of the gold sample showing significant departures from their smooth baseline LCs. Most of the undulations (${\it i.e.} \sim 76\%$) occur at post-peak phases. The energies within the undulations vary from $9.1\times10^{48}$ to $8.8\times10^{49}$\,erg, usually $<5\%$ of the integrated radiative energy. The undulation time intervals and their observed $(g-r)$ colors suggest that both the CSI and the central engine with temporal variation are possible driving mechanisms. But the central engine variation can only explain about half of the undulations while the CSI can potentially work for all undulating events. We also find that the CSM-favored fraction (62\%, 8/13) in strongly undulating events is significantly higher than that in our whole sample (23\%, 16/70), which implies the undulations tend to occur in the CSM-favored events.

\item Our careful LC modeling finds that the majority of the sample ($47/70=67\%$) can be equally well fit by both the magnetar and CSM+Ni models. This implies that LCs alone can not unambiguously identify the power mechanism for SLSNe-I. The large number of parameters in both of these models render some degeneracy which can not be broken by the LC data alone. However, a small fraction (16/70=23\%) of LCs with specific features, such as inverted V-shape, steep LC decay or the features of long rise and fast post-peak decay, are clearly much better fit by the CSM+Ni model with either wind or constant density profiles. Only 7 out of 70 LCs prefer the magnetar model. 

\item If LC undulations are indicators of CSI, our analysis and LC model fitting suggest that H-poor CSM is present in at least $25-44\%$ of the SLSN-I events. If the LCs with multiple undulations are interpreted as ejecta running into several CSM shells, this would imply that their massive progenitors experience violent, episodic mass loss events prior to the SN explosion. One such mechanism is PPISN, occurring in low-metallicity stars with ZAMS masses $>70 \,  M_\odot$ \citep{Woosley2017}. 

\end{enumerate}

We also summarize below additional statistical measurements from our sample.

\begin{enumerate}
%\item We measure \ion{Fe}{2} and \ion{O}{2} velocities and estimate the photospheric velocities of SLSNe-I in our sample. SNe\,Ic-BL and normal SNe\,Ic show a strong negative correlation between the rise times and the photospheric velocities while the correlation becomes non-significant or weak when taking into account SLSNe-I. In general, SLSNe-I have moderate velocities between those of SNe\,Ic-BL and normal SNe\,Ic.

\item The fraction of SLSNe-I with early double-peak LCs is small, about $6-44\%$ (3/15) measured from a subset of LCs with early time data. This result is consistent with that of \citet{Angus2019} based on much deeper DES data. While this feature has previously been observed only in slow-evolving events, we observe a double-peak LC in a fast-evolving SLSN-I, SN\,2019neq.

\item For the 54 events which can be fit by the magnetar model, we find $P = 2.64^{+2.58}_{-0.68}\ \mathrm{ms}$, $B_{\rm \perp} = 0.98^{+0.98}_{-0.63}\times  10^{14}\ \mathrm{G}$, $M_{\rm ej} = 5.03^{+4.01}_{-2.39}\ M_{\odot}$ and $E_{\rm k} = 2.13^{+1.89}_{-0.96} \times 10^{51}\ \mathrm{erg}$. We confirm the anti-correlation between $M_{\rm ej}$ and $P$ found previously \citep{Nicholl2017b, Blanchard2020, Hsu2021}. 

\item For the 47 events that can be fit equally well by both models, the final progenitor masses span over $6.83^{+4.04}_{-2.45}\ M_{\odot}$ and $17.92^{+24.11}_{-9.82}\ M_{\odot}$, estimated from the magnetar model and  the CSM+Ni model respectively. The CSM+Ni model thus requires a much more massive progenitor. 
    
\end{enumerate}

In conclusion, our analysis of a large number of SLSN-I LCs has revealed and confirmed several important observational properties which only become obvious after the high cadence and well sampled ZTF LCs are available. Both LC shapes and the high fraction of undulations show clear indications that CSM may be present near many SLSN-I progenitors and play critical roles in their LC energetics and evolution. Intrinsic temporal variations of the central engine can also be a possible driver for LC undulations. Our papers (I \&\ II) have put the studies of SLSN-I population on a solid statistical footing. The prospect for future progress lies with better modeling of the high quality ZTF LCs. 

\acknowledgements

We thank Dr. Weili Lin from Tsinghua University for useful discussions of CSM modeling of SLSN-I LCs. Based on observations obtained with the Samuel Oschin Telescope 48-inch and the 60-inch Telescope at the Palomar Observatory as part of the Zwicky Transient Facility project. ZTF is supported by the National Science Foundation under Grant No. AST-1440341 and a collaboration including Caltech, IPAC, the Weizmann Institute for Science, the Oskar Klein Center at Stockholm University, the University of Maryland, the University of Washington, Deutsches Elektronen-Synchrotron and Humboldt University, Los Alamos National Laboratories, the TANGO Consortium of Taiwan, the University of Wisconsin at Milwaukee, and Lawrence Berkeley National Laboratories. Operations are conducted by COO, IPAC, and UW. SED Machine is based upon work supported by the National Science Foundation under Grant No. 1106171. The ZTF forced-photometry service was funded under the Heising-Simons Foundation grant \#12540303 (PI: Graham). 

The Liverpool Telescope is operated on the island of La Palma by Liverpool John Moores University in the Spanish Observatorio del Roque de los Muchachos of the Instituto de Astrofisica de Canarias with financial support from the UK Science and Technology Facilities Council. The Nordic Optical Telescope is owned in collaboration by the University of Turku and Aarhus University, and operated jointly by Aarhus University, the University of Turku and the University of Oslo, representing Denmark, Finland and Norway, the University of Iceland and Stockholm University at the Observatorio del Roque de los Muchachos, La Palma, Spain, of the Instituto de Astrofisica de Canarias. This research has made use of data obtained through the High Energy Astrophysics Science Archive Research Center Online Service, provided by the NASA/Goddard Space Flight Center. This work was supported by the GROWTH project funded by the National Science Foundation under Grant No. 1545949. 

Z.~Chen acknowledges support from the China Scholarship Council. T.~K. acknowledges support from the Swedish National Space Agency and the Swedish Research Council. S.~S. acknowledges support from the G.R.E.A.T research environment, funded by {\em Vetenskapsr\aa det},  the Swedish Research Council, project number 2016-06012. T.-W.~C. acknowledges the EU Funding under Marie Sk\l{}odowska-Curie grant H2020-MSCA-IF-2018-842471. A.~Gal-Yam acknowledges support from the EU via ERC grant No. 725161, the ISF GW excellence center, an IMOS space infrastructure grant and BSF/Transformative and GIF grants, as well as the André Deloro Institute for Advanced Research in Space and Optics, the Schwartz/Reisman Collaborative Science Program and the Norman E Alexander Family M Foundation ULTRASAT Data Center Fund, Minerva and Yeda-Sela. R.~L. acknowledges support from a Marie Sk\l{}odowska-Curie Individual Fellowship within the Horizon 2020 European Union (EU) Framework Programme for Research and Innovation (H2020-MSCA-IF-2017-794467). The work of X.~Wang is supported by the National Natural Science Foundation of China (NSFC grants 12033003 and 11633002), the Major State Basic Research Development Program (grant 2016YFA0400803), the Scholar Program of Beijing Academy of Science and Technology (DZ:BS202002), and the Tencent XPLORER Prize.

%\facilities{PO:1.2m, PO:Hale, Liverpool:2m, NOT:2.56m, Keck:I, WHT:4.2m}

\software{
SESNspectraLib \citep{Bianco_2016}, Scikit-learn \citep{scikit-learn}, {\tt MOSFiT} \citep{Guillochon2018}}, {\tt dynesty} \citep{Speagle_2020}.

\hfill
\clearpage

\appendix

\newcounter{Atable}
\setcounter{Atable}{0}
\newcounter{Afigure}
\setcounter{Afigure}{0}

\renewcommand\thefigure{\Alph{section}\arabic{Afigure}} 
\renewcommand\thetable{\Alph{section}\arabic{Atable}} 

\section{The complete information on the ZTF SLSN-I sample}

\hfill

Table~\ref{tab:v} lists the velocities measured from spectra, as well as the spectral phase and the ionization lines.

\startlongtable
\addtocounter{Atable}{1}

\begin{deluxetable*}{lccc}
%\tabletypesize{\scriptsize}
\tablecaption{Spectral velocities.}
\tablehead{\multirow{2}{*}[-4pt]{Name} &
\colhead{Phase} &
\multirow{2}{*}[-4pt]{Ion} &
\colhead{Velocity} \\
&
(days) &
&
(${\rm km\, s^{-1}}$)
} 

\startdata
SN\,2018avk & -4.52 & \ion{Fe}{2} & $11100^{+500}_{-350}$ \\
SN\,2018don & -2.17 & \ion{Fe}{2} & $12350^{+520}_{-450}$ \\
SN\,2018don & -1.24 & \ion{Fe}{2} & $12600^{+640}_{-650}$ \\
SN\,2018don & 33.23 & \ion{Fe}{2} & $4740^{+640}_{-530}$ \\
SN\,2018bgv & 17.98 & \ion{Fe}{2} & $16790^{+180}_{-210}$ \\
SN\,2018bgv & 26.32 & \ion{Fe}{2} & $12090^{+120}_{-150}$ \\
SN\,2018lzv & 86.01 & \ion{Fe}{2} & $8180^{+740}_{-630}$ \\
SN\,2018lzv & 101.35 & \ion{Fe}{2} & $8300^{+400}_{-350}$ \\
SN\,2018lzv & 142.49 & \ion{Fe}{2} & $6700^{+600}_{-580}$ \\
\enddata
%\hline
%\bottomrule
\hspace*{\fill} \\
(This table is available in its entirety in machine-readable form.)
\label{tab:v}
\end{deluxetable*}

\hfill
\onecolumngrid
Table~\ref{tab:modeling} lists the key parameters and reduced $\chi^2$ value of the Magnetar and CSM+Ni model.

\startlongtable
\addtocounter{Atable}{1}
%\begin{deluxetable*}{p{2cm}p{1.05cm}<{\centering}p{1.05cm}<{\centering}p{1.05cm}<{\centering}p{1.05cm}<{\centering}p{1.05cm}<{\centering}p{-0.5cm}p{1.05cm}<{\centering}p{1.05cm}<{\centering}p{1.05cm}<{\centering}p{1.05cm}<{\centering}p{1.05cm}<{\centering}}

%\begin{deluxetable*}{p{1.85cm}cccccp{-0.8cm}cccccc}
\begin{deluxetable*}{lcccccccccccc}

\tabletypesize{\scriptsize}
\tablecaption{Modeling parameters}
\tablehead{
\multirow{3}{*}[-4pt]{Name} &
%\multirow{3}{*}[-4pt]{$V_{\rm spec}$} &
\multicolumn{5}{c}{Magnetar} &
\colhead{\ } &
\multicolumn{6}{c}{CSM} \\
\cline{2-6}
\cline{8-13}
& 
\colhead{$B_{\rm \perp}$} &
\colhead{$P$} &
\colhead{$M_{\rm ej}$} &
\colhead{$V_{\rm ej}$} &
\colhead{$\chi^2/dof$} & &
\colhead{$s$} &
\colhead{$M_{\rm Ni}$} &
\colhead{$M_{\rm CSM}$} &
\colhead{$M_{\rm ej}$} &
\colhead{$V_{\rm ej}$} &
\colhead{$\chi^2/dof$} \\
& 
$\rm (10^{14}G)$ &
$\rm (ms)$ &
$(M_{\odot})$ &
$(10^4 \,{\rm km\, s^{-1}})$ &
 & & &
$(M_{\odot})$ &
$(M_{\odot})$ &
$(M_{\odot})$ &
$(10^4 \,{\rm km\, s^{-1}})$ &
 } 

\startdata
SN\,2018avk & $0.16^{+0.06}_{-0.05}$ & $3.69^{+0.59}_{-0.53}$ & $2.72^{+0.76}_{-0.42}$ & $0.67^{+0.01}_{-0.00}$ & 1.38 & & 0 & $1.92^{+0.42}_{-0.38}$ & $5.49^{+1.15}_{-1.02}$ & $4.82^{+1.15}_{-0.89}$ & $0.67^{+0.01}_{-0.00}$ & 1.50 \\
SN\,2018don & $1.11^{+0.20}_{-0.23}$ & $4.60^{+0.59}_{-0.57}$ & $10.74^{+0.84}_{-0.77}$ & $0.76^{+0.00}_{-0.00}$ & 3.51 & & 2 & $3.37^{+0.27}_{-0.27}$ & $0.63^{+0.15}_{-0.17}$ & $40.78^{+3.02}_{-3.16}$ & $1.04^{+0.03}_{-0.03}$ & 2.12 \\
SN\,2018bym & $1.18^{+0.28}_{-0.21}$ & $1.92^{+0.23}_{-0.20}$ & $8.16^{+0.93}_{-0.64}$ & $1.06^{+0.02}_{-0.02}$ & 3.74 & & 0 & $3.41^{+0.19}_{-0.17}$ & $3.94^{+0.36}_{-0.32}$ & $26.69^{+2.02}_{-1.71}$ & $1.06^{+0.02}_{-0.02}$ & 3.68 \\
SN\,2018bgv & $3.03^{+0.70}_{-0.53}$ & $2.69^{+0.61}_{-0.48}$ & $1.37^{+0.54}_{-0.34}$ & $1.09^{+0.04}_{-0.03}$ & 39.30 & & 0 & $1.24^{+0.19}_{-0.09}$ & $0.20^{+0.25}_{-0.07}$ & $3.18^{+0.66}_{-0.48}$ & $1.58^{+0.04}_{-0.05}$ & 9.65 \\
SN\,2018lzv & $0.95^{+0.18}_{-0.22}$ & $0.89^{+0.28}_{-0.12}$ & $27.62^{+6.84}_{-6.35}$ & $1.04^{+0.12}_{-0.11}$ & 2.55 & & 0 & $0.35^{+0.91}_{-0.25}$ & $13.35^{+5.36}_{-3.91}$ & $52.14^{+21.31}_{-24.38}$ & $1.07^{+0.14}_{-0.14}$ & 2.58 \\
SN\,2018gbw & $1.00^{+0.42}_{-0.39}$ & $2.40^{+0.42}_{-0.52}$ & $4.36^{+1.85}_{-1.11}$ & $1.03^{+0.07}_{-0.06}$ & 2.71 & & 2 & $0.10^{+1.51}_{-0.09}$ & $6.06^{+2.76}_{-1.44}$ & $8.02^{+28.67}_{-6.52}$ & $1.22^{+0.11}_{-0.12}$ & 2.84 \\
SN\,2018fcg & $4.92^{+0.65}_{-1.20}$ & $6.77^{+1.23}_{-1.94}$ & $2.73^{+0.76}_{-0.65}$ & $0.91^{+0.03}_{-0.03}$ & 1.72 & & 2 & $0.58^{+0.06}_{-0.12}$ & $1.97^{+0.38}_{-0.68}$ & $1.91^{+0.91}_{-0.60}$ & $0.95^{+0.12}_{-0.04}$ & 2.53 \\
SN\,2018gft & $0.17^{+0.04}_{-0.04}$ & $2.00^{+0.23}_{-0.23}$ & $7.30^{+0.84}_{-0.65}$ & $1.06^{+0.03}_{-0.03}$ & 5.16 & & 2 & $0.01^{+0.03}_{-0.01}$ & $18.83^{+2.74}_{-2.07}$ & $0.22^{+0.13}_{-0.08}$ & $1.28^{+0.04}_{-0.03}$ & 10.39 \\
SN\,2018lzx & $0.28^{+0.06}_{-0.06}$ & $2.09^{+0.21}_{-0.23}$ & $19.00^{+3.81}_{-3.12}$ & $0.51^{+0.02}_{-0.02}$ & 1.40 & & 0 & $24.35^{+3.15}_{-4.04}$ & $22.30^{+5.25}_{-3.83}$ & $95.95^{+2.84}_{-6.32}$ & $0.63^{+0.03}_{-0.02}$ & 1.80 \\
SN\,2018hpq & $1.58^{+0.50}_{-0.33}$ & $6.93^{+0.88}_{-0.88}$ & $4.29^{+2.12}_{-1.35}$ & $0.55^{+0.04}_{-0.04}$ & 2.52 & & 0 & $0.01^{+0.02}_{-0.00}$ & $9.84^{+3.39}_{-3.00}$ & $1.35^{+4.12}_{-0.92}$ & $0.54^{+0.03}_{-0.03}$ & 1.69 \\
SN\,2018lfe & $1.63^{+0.45}_{-0.40}$ & $3.96^{+0.55}_{-0.60}$ & $2.31^{+0.94}_{-0.67}$ & $0.98^{+0.10}_{-0.10}$ & 0.54 & & 2 & $1.51^{+2.03}_{-1.36}$ & $1.63^{+0.89}_{-0.42}$ & $19.64^{+21.98}_{-14.89}$ & $1.35^{+0.15}_{-0.12}$ & 1.25 \\
SN\,2018hti & $0.35^{+0.06}_{-0.07}$ & $2.84^{+0.17}_{-0.28}$ & $4.04^{+0.91}_{-0.61}$ & $0.84^{+0.01}_{-0.00}$ & 12.30 & & 0 & $3.10^{+0.51}_{-0.53}$ & $18.39^{+3.31}_{-3.02}$ & $24.62^{+14.31}_{-10.50}$ & $0.92^{+0.02}_{-0.02}$ & 18.77 \\
SN\,2018lfd & $1.26^{+0.31}_{-0.24}$ & $1.43^{+0.33}_{-0.30}$ & $9.10^{+4.15}_{-2.51}$ & $0.85^{+0.08}_{-0.07}$ & 1.17 & & 0 & $2.99^{+0.62}_{-0.49}$ & $5.41^{+3.49}_{-1.92}$ & $49.08^{+16.06}_{-12.08}$ & $0.92^{+0.09}_{-0.10}$ & 1.16 \\
SN\,2018kyt & $1.32^{+0.22}_{-0.28}$ & $9.05^{+0.73}_{-1.01}$ & $2.03^{+1.05}_{-0.61}$ & $0.65^{+0.03}_{-0.02}$ & 6.36 & & 2 & $2.30^{+0.14}_{-0.10}$ & $1.51^{+0.27}_{-0.23}$ & $11.92^{+1.01}_{-1.23}$ & $0.66^{+0.03}_{-0.02}$ & 4.03 \\
SN\,2019J & $3.33^{+1.01}_{-0.80}$ & $1.42^{+0.86}_{-0.50}$ & $10.60^{+5.57}_{-3.31}$ & $0.50^{+0.05}_{-0.05}$ & 3.75 & & 0 & $0.18^{+0.18}_{-0.10}$ & $6.26^{+2.10}_{-1.88}$ & $33.42^{+10.78}_{-8.57}$ & $0.43^{+0.04}_{-0.03}$ & 2.84 \\
SN\,2019cca & $1.00^{+0.32}_{-0.25}$ & $2.17^{+0.57}_{-0.70}$ & $6.03^{+3.67}_{-2.21}$ & $0.87^{+0.09}_{-0.08}$ & 0.77 & & 0 & $0.80^{+2.60}_{-0.67}$ & $10.96^{+5.73}_{-6.23}$ & $28.16^{+24.14}_{-12.11}$ & $0.85^{+0.15}_{-0.12}$ & 0.84 \\
SN\,2019bgu & $4.30^{+1.07}_{-0.97}$ & $2.30^{+0.71}_{-0.66}$ & $5.21^{+1.52}_{-0.98}$ & $1.05^{+0.04}_{-0.02}$ & 1.74 & & 2 & $0.26^{+0.26}_{-0.23}$ & $1.42^{+0.60}_{-0.32}$ & $2.01^{+9.26}_{-1.52}$ & $1.09^{+0.06}_{-0.05}$ & 1.89 \\
SN\,2019kwq & $0.43^{+0.09}_{-0.09}$ & $2.63^{+0.27}_{-0.28}$ & $6.73^{+2.04}_{-1.47}$ & $0.70^{+0.04}_{-0.03}$ & 1.19 & & 0 & $22.04^{+1.50}_{-1.08}$ & $11.47^{+1.95}_{-1.76}$ & $78.56^{+9.65}_{-9.29}$ & $0.71^{+0.03}_{-0.03}$ & 1.06 \\
SN\,2019dgr & $0.98^{+0.34}_{-0.26}$ & $2.64^{+0.71}_{-0.79}$ & $4.26^{+1.80}_{-1.25}$ & $1.05^{+0.10}_{-0.09}$ & 1.79 & & 0 & $0.10^{+0.48}_{-0.07}$ & $4.00^{+1.17}_{-0.71}$ & $14.44^{+4.18}_{-3.90}$ & $1.13^{+0.10}_{-0.08}$ & 1.66 \\
SN\,2019kws & $1.76^{+0.20}_{-0.44}$ & $8.32^{+0.56}_{-1.27}$ & $2.76^{+0.73}_{-0.54}$ & $0.73^{+0.03}_{-0.03}$ & 1.62 & & 2 & $0.02^{+1.36}_{-0.02}$ & $3.75^{+0.36}_{-0.99}$ & $0.17^{+3.54}_{-0.05}$ & $0.78^{+0.03}_{-0.06}$ & 1.49 \\
SN\,2019cdt & $2.95^{+0.94}_{-0.85}$ & $1.49^{+0.58}_{-0.44}$ & $5.27^{+1.88}_{-1.34}$ & $1.55^{+0.13}_{-0.13}$ & 7.57 & & 2 & $0.00^{+0.06}_{-0.00}$ & $2.52^{+0.59}_{-0.24}$ & $0.59^{+17.33}_{-0.40}$ & $1.55^{+0.08}_{-0.12}$ & 4.57 \\
SN\,2019aamp & $0.41^{+0.13}_{-0.13}$ & $2.77^{+0.41}_{-0.32}$ & $2.73^{+0.80}_{-0.62}$ & $1.16^{+0.07}_{-0.12}$ & 0.84 & & 0 & $0.12^{+0.79}_{-0.09}$ & $6.11^{+2.37}_{-2.91}$ & $4.61^{+12.50}_{-2.01}$ & $1.36^{+0.06}_{-0.09}$ & 0.96 \\
SN\,2019dlr & $1.48^{+0.60}_{-0.29}$ & $4.13^{+0.66}_{-1.44}$ & $5.37^{+3.03}_{-1.49}$ & $0.73^{+0.06}_{-0.06}$ & 2.01 & & 0 & $0.64^{+1.18}_{-0.51}$ & $3.24^{+0.43}_{-0.43}$ & $32.49^{+12.05}_{-5.99}$ & $0.76^{+0.04}_{-0.05}$ & 1.95 \\
SN\,2019cwu & $0.80^{+0.15}_{-0.16}$ & $4.80^{+0.46}_{-0.44}$ & $1.17^{+0.34}_{-0.20}$ & $1.15^{+0.03}_{-0.03}$ & 1.09 & & 0 & $0.11^{+0.27}_{-0.10}$ & $3.03^{+1.97}_{-0.42}$ & $2.17^{+1.12}_{-1.63}$ & $1.28^{+0.06}_{-0.05}$ & 1.12 \\
SN\,2019kwt & $0.20^{+0.05}_{-0.04}$ & $1.80^{+0.29}_{-0.26}$ & $12.01^{+1.58}_{-1.40}$ & $0.83^{+0.04}_{-0.04}$ & 2.02 & & 0 & $0.30^{+1.96}_{-0.26}$ & $20.02^{+5.63}_{-3.42}$ & $20.69^{+6.62}_{-14.30}$ & $0.73^{+0.04}_{-0.04}$ & 0.79 \\
SN\,2019eot & $0.67^{+0.18}_{-0.15}$ & $2.40^{+0.27}_{-0.25}$ & $7.70^{+0.84}_{-0.75}$ & $0.98^{+0.02}_{-0.02}$ & 4.27 & & 0 & $0.30^{+0.69}_{-0.22}$ & $4.55^{+0.43}_{-0.38}$ & $26.31^{+2.43}_{-1.84}$ & $1.09^{+0.02}_{-0.02}$ & 12.38 \\
SN\,2019kwu & $1.09^{+0.31}_{-0.29}$ & $2.53^{+0.41}_{-0.47}$ & $2.90^{+1.17}_{-0.72}$ & $1.14^{+0.03}_{-0.05}$ & 0.61 & & 0 & $0.18^{+1.11}_{-0.15}$ & $6.28^{+1.71}_{-1.10}$ & $16.59^{+3.80}_{-5.49}$ & $1.15^{+0.03}_{-0.04}$ & 0.65 \\
SN\,2019gqi & $0.73^{+0.21}_{-0.23}$ & $3.70^{+0.76}_{-0.94}$ & $2.72^{+0.94}_{-0.70}$ & $1.07^{+0.07}_{-0.07}$ & 0.66 & & 0 & $0.52^{+1.98}_{-0.49}$ & $3.38^{+3.05}_{-1.89}$ & $11.37^{+14.66}_{-9.43}$ & $1.14^{+0.13}_{-0.08}$ & 0.64 \\
SN\,2019fiy & $0.47^{+0.39}_{-0.28}$ & $1.68^{+0.41}_{-0.48}$ & $3.96^{+2.80}_{-1.61}$ & $1.29^{+0.19}_{-0.18}$ & 2.95 & & 0 & $0.05^{+0.32}_{-0.04}$ & $6.60^{+2.95}_{-2.00}$ & $2.71^{+4.90}_{-1.87}$ & $2.24^{+0.17}_{-0.29}$ & 3.76 \\
SN\,2019gam & $0.88^{+0.38}_{-0.43}$ & $2.70^{+3.44}_{-1.52}$ & $14.16^{+13.53}_{-7.60}$ & $0.55^{+0.10}_{-0.09}$ & 1.86 & & 0 & $0.03^{+0.12}_{-0.02}$ & $14.36^{+3.82}_{-3.07}$ & $2.17^{+3.51}_{-1.23}$ & $0.70^{+0.09}_{-0.08}$ & 2.08 \\
SN\,2019gfm & $1.97^{+0.43}_{-0.40}$ & $6.54^{+0.67}_{-0.80}$ & $1.55^{+0.49}_{-0.38}$ & $1.07^{+0.04}_{-0.03}$ & 2.15 & & 0 & $0.02^{+0.17}_{-0.02}$ & $2.16^{+0.67}_{-0.49}$ & $0.89^{+1.04}_{-0.53}$ & $1.16^{+0.06}_{-0.08}$ & 2.92 \\
SN\,2019hge & $2.21^{+0.37}_{-0.48}$ & $7.42^{+0.71}_{-0.84}$ & $8.98^{+0.98}_{-0.81}$ & $0.66^{+0.01}_{-0.00}$ & 9.03 & & 2 & $0.00^{+0.01}_{-0.00}$ & $6.20^{+0.50}_{-0.39}$ & $0.12^{+0.02}_{-0.01}$ & $0.66^{+0.01}_{-0.00}$ & 6.65 \\
SN\,2019hno & $2.06^{+0.37}_{-0.44}$ & $4.81^{+0.53}_{-0.62}$ & $3.26^{+0.60}_{-0.44}$ & $0.94^{+0.04}_{-0.04}$ & 1.41 & & 0 & $2.13^{+0.30}_{-0.27}$ & $2.01^{+0.29}_{-0.23}$ & $8.58^{+1.02}_{-0.75}$ & $0.88^{+0.03}_{-0.03}$ & 1.09 \\
SN\,2019aamq & $0.22^{+0.05}_{-0.05}$ & $2.02^{+0.39}_{-0.33}$ & $8.47^{+1.64}_{-1.35}$ & $0.86^{+0.02}_{-0.01}$ & 1.60 & & 2 & $0.57^{+2.68}_{-0.48}$ & $22.39^{+4.09}_{-3.81}$ & $9.45^{+39.94}_{-7.54}$ & $0.87^{+0.03}_{-0.02}$ & 1.46 \\
SN\,2019kcy & $0.35^{+0.14}_{-0.11}$ & $2.43^{+0.29}_{-0.28}$ & $5.10^{+1.48}_{-1.22}$ & $0.99^{+0.05}_{-0.05}$ & 0.57 & & 2 & $0.01^{+0.05}_{-0.01}$ & $13.01^{+3.71}_{-2.76}$ & $0.54^{+1.03}_{-0.32}$ & $1.36^{+0.08}_{-0.07}$ & 1.00 \\
SN\,2019aamx & $0.81^{+0.38}_{-0.29}$ & $2.54^{+0.80}_{-0.77}$ & $9.83^{+3.85}_{-3.10}$ & $0.76^{+0.07}_{-0.05}$ & 0.58 & & 0 & $1.67^{+3.18}_{-1.19}$ & $4.04^{+1.21}_{-0.69}$ & $57.51^{+11.87}_{-10.94}$ & $0.78^{+0.04}_{-0.04}$ & 0.39 \\
SN\,2019aamr & $1.07^{+0.42}_{-0.37}$ & $4.17^{+0.85}_{-1.13}$ & $2.22^{+1.13}_{-0.58}$ & $1.19^{+0.10}_{-0.10}$ & 0.73 & & 2 & $0.02^{+0.09}_{-0.01}$ & $3.61^{+1.30}_{-0.60}$ & $0.30^{+0.53}_{-0.16}$ & $1.38^{+0.12}_{-0.10}$ & 0.71 \\
SN\,2019lsq & $0.97^{+0.19}_{-0.20}$ & $5.98^{+0.58}_{-0.64}$ & $2.67^{+0.74}_{-0.44}$ & $0.85^{+0.01}_{-0.00}$ & 7.50 & & 0 & $0.03^{+0.13}_{-0.02}$ & $6.39^{+1.79}_{-0.96}$ & $1.00^{+0.58}_{-0.54}$ & $0.86^{+0.01}_{-0.01}$ & 9.38 \\
SN\,2019nhs & $1.49^{+0.35}_{-0.29}$ & $3.82^{+0.43}_{-0.47}$ & $3.91^{+0.70}_{-0.47}$ & $0.94^{+0.02}_{-0.01}$ & 3.28 & & 2 & $0.36^{+0.30}_{-0.30}$ & $3.60^{+0.44}_{-0.27}$ & $19.23^{+4.38}_{-16.27}$ & $0.95^{+0.05}_{-0.02}$ & 1.70 \\
SN\,2019aams & $1.38^{+0.44}_{-0.35}$ & $2.05^{+0.57}_{-0.62}$ & $4.96^{+2.32}_{-1.37}$ & $1.19^{+0.12}_{-0.10}$ & 1.09 & & 0 & $1.56^{+4.40}_{-1.41}$ & $2.42^{+4.38}_{-0.67}$ & $16.66^{+9.72}_{-13.88}$ & $1.32^{+0.30}_{-0.10}$ & 0.94 \\
SN\,2019neq & $0.43^{+0.08}_{-0.08}$ & $3.72^{+0.33}_{-0.38}$ & $1.74^{+0.29}_{-0.22}$ & $1.13^{+0.03}_{-0.03}$ & 13.07 & & 0 & $0.56^{+1.13}_{-0.22}$ & $2.03^{+0.18}_{-0.28}$ & $1.29^{+2.45}_{-0.49}$ & $1.29^{+0.05}_{-0.05}$ & 14.31 \\
SN\,2019sgg & $1.18^{+0.29}_{-0.26}$ & $0.99^{+0.28}_{-0.19}$ & $17.09^{+7.58}_{-3.95}$ & $1.02^{+0.03}_{-0.02}$ & 1.51 & & 0 & $1.99^{+8.22}_{-1.77}$ & $4.21^{+0.69}_{-0.62}$ & $56.15^{+6.22}_{-5.49}$ & $1.03^{+0.03}_{-0.02}$ & 1.04 \\
SN\,2019aamt & $1.17^{+0.30}_{-0.25}$ & $5.60^{+0.53}_{-0.63}$ & $1.59^{+1.03}_{-0.63}$ & $0.67^{+0.07}_{-0.07}$ & 1.07 & & 0 & $0.20^{+1.34}_{-0.17}$ & $4.41^{+1.76}_{-1.52}$ & $12.10^{+6.30}_{-5.36}$ & $0.75^{+0.07}_{-0.06}$ & 1.05 \\
SN\,2019sgh & $1.78^{+0.38}_{-0.32}$ & $2.14^{+1.51}_{-0.70}$ & $2.62^{+0.95}_{-0.62}$ & $0.97^{+0.07}_{-0.06}$ & 1.43 & & 0 & $0.05^{+0.18}_{-0.04}$ & $2.41^{+0.92}_{-0.51}$ & $6.54^{+2.47}_{-2.65}$ & $1.18^{+0.09}_{-0.11}$ & 1.60 \\
SN\,2019stc & $0.95^{+0.19}_{-0.17}$ & $6.76^{+1.25}_{-1.03}$ & $3.13^{+0.84}_{-0.72}$ & $0.88^{+0.08}_{-0.07}$ & 2.25 & & 0 & $5.70^{+0.66}_{-0.57}$ & $3.83^{+1.72}_{-1.17}$ & $25.88^{+8.89}_{-6.87}$ & $0.75^{+0.06}_{-0.05}$ & 1.24 \\
SN\,2019szu & $0.79^{+0.20}_{-0.20}$ & $1.40^{+0.60}_{-0.39}$ & $47.92^{+13.16}_{-12.98}$ & $0.59^{+0.08}_{-0.07}$ & 2.58 & & 0 & $9.30^{+3.27}_{-3.65}$ & $15.79^{+6.52}_{-4.07}$ & $70.43^{+18.78}_{-22.03}$ & $0.52^{+0.04}_{-0.03}$ & 4.73 \\
SN\,2019unb & $1.99^{+0.41}_{-0.42}$ & $3.62^{+1.05}_{-0.80}$ & $8.23^{+3.33}_{-1.93}$ & $0.43^{+0.03}_{-0.03}$ & 24.62 & & 0 & $1.06^{+0.20}_{-0.12}$ & $2.48^{+1.19}_{-0.62}$ & $36.57^{+12.57}_{-6.67}$ & $0.46^{+0.03}_{-0.03}$ & 14.40 \\
SN\,2019ujb & $2.26^{+0.53}_{-0.52}$ & $0.96^{+0.25}_{-0.17}$ & $21.09^{+7.08}_{-4.60}$ & $1.07^{+0.05}_{-0.03}$ & 4.77 & & 0 & $0.05^{+0.11}_{-0.04}$ & $5.31^{+0.66}_{-0.56}$ & $0.69^{+0.40}_{-0.22}$ & $1.05^{+0.03}_{-0.02}$ & 2.25 \\
SN\,2019xdy & $2.02^{+0.70}_{-0.42}$ & $3.04^{+1.35}_{-1.39}$ & $6.27^{+4.02}_{-2.12}$ & $0.75^{+0.09}_{-0.08}$ & 1.16 & & 0 & $0.01^{+0.02}_{-0.00}$ & $9.68^{+2.62}_{-3.12}$ & $1.48^{+2.41}_{-0.96}$ & $0.67^{+0.09}_{-0.08}$ & 1.06 \\
SN\,2019aamw & $0.75^{+0.27}_{-0.20}$ & $5.70^{+0.56}_{-0.71}$ & $5.48^{+2.29}_{-1.54}$ & $0.56^{+0.05}_{-0.05}$ & 1.09 & & 0 & $0.15^{+1.17}_{-0.13}$ & $14.37^{+6.92}_{-5.61}$ & $5.19^{+10.49}_{-3.54}$ & $0.54^{+0.06}_{-0.05}$ & 0.95 \\
SN\,2019zbv & $0.32^{+0.18}_{-0.08}$ & $2.59^{+0.26}_{-0.20}$ & $4.51^{+1.48}_{-0.84}$ & $0.92^{+0.04}_{-0.03}$ & 0.87 & & 0 & $0.34^{+0.51}_{-0.19}$ & $14.50^{+2.60}_{-2.51}$ & $23.31^{+8.99}_{-10.99}$ & $0.84^{+0.05}_{-0.04}$ & 0.92 \\
SN\,2020fvm & $0.10^{+0.02}_{-0.02}$ & $2.08^{+0.21}_{-0.24}$ & $21.32^{+1.08}_{-1.14}$ & $0.67^{+0.03}_{-0.02}$ & 10.61 & & 0 & $14.84^{+0.42}_{-0.42}$ & $6.45^{+0.28}_{-0.22}$ & $32.23^{+2.18}_{-1.43}$ & $0.65^{+0.01}_{-0.01}$ & 4.82 \\
SN\,2019aamv & $1.45^{+0.28}_{-0.33}$ & $0.93^{+0.22}_{-0.15}$ & $40.37^{+16.66}_{-8.49}$ & $0.75^{+0.01}_{-0.00}$ & 2.18 & & 0 & $14.50^{+0.46}_{-0.36}$ & $1.96^{+0.11}_{-0.09}$ & $30.69^{+2.22}_{-1.38}$ & $0.75^{+0.00}_{-0.00}$ & 1.32 \\
SN\,2020ank & $2.33^{+0.48}_{-0.47}$ & $1.62^{+0.45}_{-0.45}$ & $5.44^{+2.04}_{-1.10}$ & $1.33^{+0.08}_{-0.06}$ & 1.98 & & 0 & $0.04^{+0.05}_{-0.02}$ & $2.11^{+1.04}_{-0.65}$ & $7.52^{+4.35}_{-2.28}$ & $1.30^{+0.07}_{-0.04}$ & 1.83 \\
SN\,2020aup & $0.86^{+0.76}_{-0.33}$ & $3.83^{+0.47}_{-0.60}$ & $1.97^{+0.70}_{-0.47}$ & $1.01^{+0.08}_{-0.08}$ & 1.44 & & 0 & $0.87^{+1.32}_{-0.81}$ & $2.06^{+1.20}_{-0.79}$ & $10.09^{+19.64}_{-2.37}$ & $1.07^{+0.10}_{-0.12}$ & 1.74 \\
SN\,2020auv & $0.80^{+0.17}_{-0.16}$ & $3.64^{+0.48}_{-0.48}$ & $2.40^{+1.19}_{-0.69}$ & $1.03^{+0.14}_{-0.15}$ & 6.32 & & 2 & $6.59^{+0.37}_{-0.29}$ & $5.07^{+1.33}_{-0.98}$ & $27.05^{+7.09}_{-5.17}$ & $1.00^{+0.06}_{-0.07}$ & 1.44 \\
SN\,2020dlb & $0.15^{+0.04}_{-0.03}$ & $1.54^{+0.19}_{-0.18}$ & $6.76^{+0.79}_{-0.84}$ & $1.46^{+0.02}_{-0.01}$ & 4.79 & & 2 & $0.02^{+0.21}_{-0.02}$ & $12.72^{+3.06}_{-3.26}$ & $0.75^{+2.34}_{-0.46}$ & $1.48^{+0.04}_{-0.02}$ & 2.59 \\
SN\,2020fyq & $1.94^{+0.53}_{-0.52}$ & $4.54^{+1.11}_{-1.85}$ & $18.34^{+9.79}_{-5.67}$ & $0.41^{+0.02}_{-0.03}$ & 1.35 & & 0 & $0.13^{+0.27}_{-0.07}$ & $7.85^{+1.50}_{-1.47}$ & $44.23^{+6.50}_{-4.47}$ & $0.35^{+0.01}_{-0.01}$ & 1.24 \\
SN\,2020exj & $4.44^{+0.71}_{-0.78}$ & $6.44^{+0.70}_{-0.79}$ & $2.51^{+1.25}_{-0.80}$ & $0.65^{+0.02}_{-0.02}$ & 1.67 & & 2 & $0.04^{+0.04}_{-0.02}$ & $1.62^{+0.25}_{-0.20}$ & $21.53^{+2.49}_{-2.38}$ & $0.70^{+0.03}_{-0.03}$ & 2.11 \\
SN\,2020htd & $0.41^{+0.08}_{-0.09}$ & $3.79^{+0.35}_{-0.43}$ & $6.69^{+2.93}_{-1.81}$ & $0.52^{+0.05}_{-0.05}$ & 5.21 & & 0 & $3.05^{+1.10}_{-1.00}$ & $10.89^{+1.28}_{-1.27}$ & $94.86^{+3.43}_{-5.92}$ & $0.48^{+0.02}_{-0.02}$ & 1.71 \\
SN\,2020iyj & $0.66^{+0.17}_{-0.20}$ & $3.21^{+0.37}_{-0.67}$ & $4.97^{+1.33}_{-1.67}$ & $0.90^{+0.06}_{-0.06}$ & 1.26 & & 0 & $1.09^{+1.56}_{-0.76}$ & $9.87^{+3.40}_{-3.84}$ & $15.68^{+12.34}_{-8.28}$ & $0.91^{+0.11}_{-0.08}$ & 1.28 \\
SN\,2020kox & $0.22^{+0.05}_{-0.05}$ & $2.58^{+0.27}_{-0.29}$ & $3.81^{+1.17}_{-0.87}$ & $0.93^{+0.10}_{-0.05}$ & 2.25 & & 0 & $0.56^{+11.36}_{-0.50}$ & $19.06^{+5.76}_{-10.70}$ & $22.10^{+34.95}_{-15.16}$ & $0.97^{+0.16}_{-0.08}$ & 2.99 \\
SN\,2020jii & $1.06^{+0.23}_{-0.21}$ & $2.63^{+0.40}_{-0.41}$ & $5.43^{+1.27}_{-0.89}$ & $0.97^{+0.04}_{-0.05}$ & 1.35 & & 0 & $6.61^{+1.24}_{-1.31}$ & $2.82^{+0.45}_{-0.38}$ & $21.65^{+3.46}_{-2.61}$ & $1.09^{+0.04}_{-0.05}$ & 1.25 \\
SN\,2020afah & $0.07^{+0.03}_{-0.02}$ & $1.58^{+0.26}_{-0.20}$ & $3.13^{+0.43}_{-0.32}$ & $0.88^{+0.04}_{-0.04}$ & 1.15 & & 0 & $7.18^{+2.34}_{-2.64}$ & $11.02^{+3.86}_{-3.10}$ & $20.65^{+6.49}_{-4.20}$ & $0.98^{+0.05}_{-0.06}$ & 1.03 \\
SN\,2020afag & $0.85^{+0.13}_{-0.16}$ & $3.58^{+0.34}_{-0.38}$ & $4.82^{+1.01}_{-0.76}$ & $0.84^{+0.04}_{-0.04}$ & 1.98 & & 0 & $12.60^{+1.12}_{-1.03}$ & $4.67^{+1.12}_{-1.06}$ & $42.84^{+7.68}_{-7.55}$ & $0.87^{+0.04}_{-0.04}$ & 1.21 \\
SN\,2020onb & $0.53^{+0.15}_{-0.12}$ & $4.27^{+0.47}_{-0.48}$ & $5.67^{+0.85}_{-0.77}$ & $0.83^{+0.02}_{-0.01}$ & 2.50 & & 0 & $0.04^{+0.16}_{-0.03}$ & $8.39^{+2.51}_{-1.75}$ & $0.77^{+0.87}_{-0.43}$ & $0.84^{+0.03}_{-0.02}$ & 2.46 \\
SN\,2020qef & $1.13^{+0.22}_{-0.24}$ & $8.45^{+0.77}_{-0.94}$ & $2.82^{+1.26}_{-0.77}$ & $0.65^{+0.04}_{-0.04}$ & 4.12 & & 2 & $3.16^{+0.18}_{-0.18}$ & $2.23^{+0.29}_{-0.27}$ & $10.49^{+1.40}_{-1.18}$ & $0.61^{+0.02}_{-0.02}$ & 0.80 \\
SN\,2020rmv & $0.34^{+0.07}_{-0.07}$ & $3.14^{+0.32}_{-0.33}$ & $5.32^{+2.20}_{-1.33}$ & $0.68^{+0.03}_{-0.03}$ & 3.19 & & 2 & $0.02^{+0.13}_{-0.01}$ & $16.71^{+4.04}_{-3.41}$ & $1.70^{+3.08}_{-1.01}$ & $0.90^{+0.05}_{-0.04}$ & 4.82 \\
SN\,2020xkv & $0.09^{+0.03}_{-0.03}$ & $2.65^{+0.45}_{-0.50}$ & $8.09^{+2.08}_{-1.00}$ & $0.89^{+0.03}_{-0.03}$ & 1.04 & & 0 & $0.00^{+0.03}_{-0.00}$ & $29.43^{+0.39}_{-0.73}$ & $0.14^{+0.05}_{-0.03}$ & $0.82^{+0.02}_{-0.03}$ & 1.11 \\
SN\,2020xgd & $1.66^{+0.65}_{-0.46}$ & $2.11^{+0.74}_{-0.69}$ & $5.33^{+2.27}_{-1.30}$ & $1.14^{+0.09}_{-0.07}$ & 1.14 & & 2 & $1.13^{+3.77}_{-1.10}$ & $3.89^{+1.58}_{-0.65}$ & $7.31^{+12.12}_{-5.90}$ & $1.16^{+0.08}_{-0.07}$ & 0.93 \\
\enddata
\label{tab:modeling}
\end{deluxetable*}

\begin{figure*}[htp]
\addtocounter{Afigure}{1}
\centering
\includegraphics[width=\textwidth]{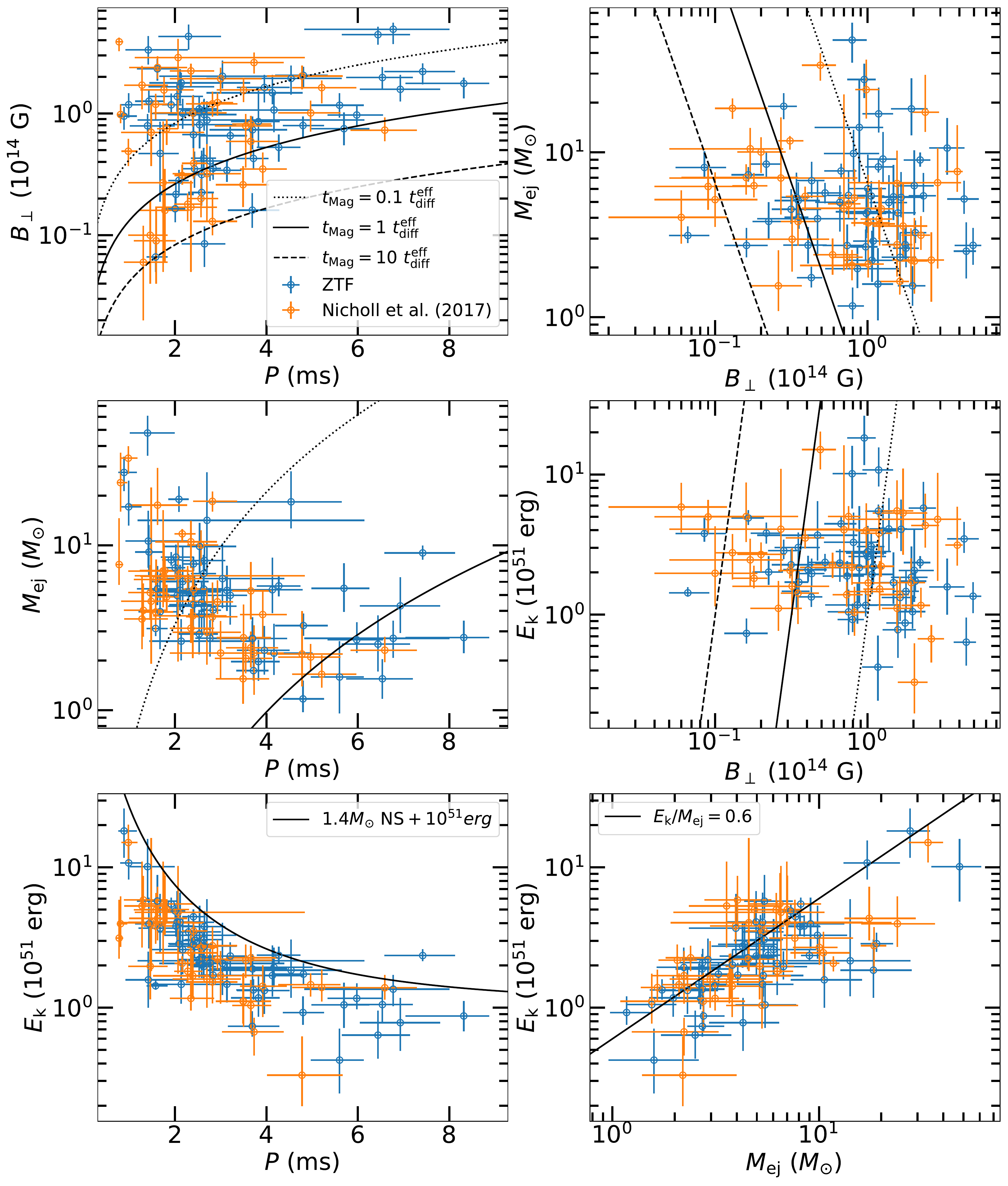}
\caption{Values and 1$\sigma$ errors of key parameters ($P$, $B_{\rm \perp}$, $M_{\rm ej}$, $E_{\rm k}$) for the 7 SLSNe-I favored by magnetar models and the 47 equally well-fit ones. The values from our sample are marked in blue while those from \citet{Nicholl2017b} are marked in orange. $t_{\rm mag}$ is the spin-down time scale for the magnetar model and the ratio of $t_{\rm mag}/t_{\rm diff}^{\rm eff}$ follows 
%the Equation~11 in \citet{Nicholl2017b}
\citet[][their equation 11]{Nicholl2017b}.
%The spin down time scale for magnetar model can be calculated by $t_{\rm mag} \simeq 1.5 \times \left( \frac{M_{\rm NS}}{1.4M_{\odot}} \right)^{3/2} \left( \frac{P}{1 \mathrm{ms}} \right)^2 \left( \frac{B_{\rm \perp}}{10^{14} \mathrm{G}} \right)^{-2}$\,days.
}
\label{fig:mosfitcompare}
\end{figure*}

\hfill
\onecolumngrid
In Figure~\ref{fig:mosfitcompare}, we compare the key parameters of the magnetar model with those from \citet{Nicholl2017b}.

%% For this sample we use BibTeX plus aasjournals.bst to generate the
%% the bibliography. The sample63.bib file was populated from ADS. To
%% get the citations to show in the compiled file do the following:
%%
%% pdflatex sample63.tex
%% bibtext sample63
%% pdflatex sample63.tex
%% pdflatex sample63.tex

\clearpage
\bibliography{draft.bib}
\bibliographystyle{aasjournal}

%% This command is needed to show the entire author+affiliation list when
%% the collaboration and author truncation commands are used.  It has to
%% go at the end of the manuscript.
%\allauthors

%% Include this line if you are using the \added, \replaced, \deleted
%% commands to see a summary list of all changes at the end of the article.
%\listofchanges

\end{document}